\documentclass[12pt]{article}
\usepackage[english]{babel}
\usepackage{amsmath,amssymb,dsfont, float, bm}
\usepackage{amsmath,amssymb,amsfonts,amsthm,mathtools,mathrsfs,booktabs}
\usepackage{multicol}
\usepackage{multirow}
\usepackage{tikz,float}
\usepackage{pgfplots}
\pgfplotsset{compat=1.16}
\usepackage{authblk}
\usepackage{caption}
\usepackage{graphicx}
\usepackage{subcaption}
\usepackage{rotating}

\usepackage{natbib}

\usepackage[colorlinks=true, citecolor=blue, urlcolor=blue]{hyperref}
\usepackage{newtxtext,newtxmath} 
\usepackage{multirow}
\usepackage{bookmark}
\usepackage{booktabs}
\usepackage{makecell}
\usepackage{setspace}
\hypersetup{                 
	colorlinks=true,
	linkcolor=blue,
	citecolor=blue,
	anchorcolor=blue,
	urlcolor=blue,
	linktoc=page,
}
\usepackage{pdflscape}
\usepackage{adjustbox}
\usepackage{chngcntr}
\usepackage{threeparttable}
\usepackage{caption}

\usepackage[a4paper,margin=2.5cm]{geometry}

\usepackage{multirow}
\hypersetup{                 
	colorlinks=true,
	linkcolor=blue,
	citecolor=blue,
	anchorcolor=blue,
	urlcolor=blue,
	linktoc=page,
}

\begin{document}
	
	\title{\Large \bf A parametric mixture cure rate model with competing risks}
	
	\author[1]{Carolyne Soares de Brito}
	\author[1]{\, Frederico Machado
		Almeida \thanks{Corresponding author: frederico.almeida@unb.br}}
	\affil[1]{Department of Statistics, University of Bras\'ilia, 70910-900, Bras\'ilia, Brazil.}
	\setcounter{Maxaffil}{0}

	\maketitle
	\section*{Abstract}
	
	The well-known standard mixture cure fraction model has become a widely adopted approach for modeling long-term survivors, as it accounts for both cured and non-cured individuals in the study population. Since its seminal introduction by Berkson and Gage, numerous extensions have been proposed to enhance the flexibility and applicability of the model across diverse clinical settings. However, most of these developments are constrained by their exclusive focus on a single cause of failure within the non-cured subgroup, often treating events from alternative causes as censored. Such a restriction may fail to capture the underlying heterogeneity of real-world survival data, where individuals are frequently exposed to multiple, competing causes of failure. In these contexts, traditional modeling approaches often lack the sophistication needed to capture the intricate nature of survival outcomes, frequently yielding unreliable estimates of cause-specific event probabilities. Moreover, treating competing risks as non-informative censoring can exacerbate bias in parameter estimation and lead to misinterpretation in clinical settings. These methodological shortcomings highlight the need for advanced, flexible survival modeling frameworks that can seamlessly integrate multiple competing risks while accounting for long-term survivors. Motivated by these considerations, we investigate an extension of the standard mixture cure model that simultaneously accounts for a cure fraction and multiple competing causes of failure. Model parameters are estimated using a maximum likelihood approach applied to augmented data. The estimators’ performance is rigorously assessed through comprehensive simulation studies, which demonstrate robust estimation across the models considered, with key metrics consistently falling within acceptable bounds. Finally, we demonstrate the applicability of the proposed methods using melanoma patient data from the Surveillance, Epidemiology, and End Results (SEER) database, which includes both long-term survivors and multiple competing causes of failure.

	\vspace{0.3cm}
	\noindent {\em Keywords:} augmented data, competing risks, mixture models, parametric approach, SEER database.
	
	\section{Introduction}\label{sec1}
	
	Standard survival models assume that all individuals are at risk of experiencing the event of interest. In healthcare, however, recent technological advancements, particularly in detection methods and treatments associated with early prognosis, have made it increasingly common to observe individuals who, by the end of a study, exhibit a favorable response to treatment. Such individuals are commonly referred to as long-term survivors, cured, or non-susceptible, and their survival times are typically assumed to be infinite \citep{maller1994testing,masud2018variable}.
	
	Cure fraction models have been extensively studied in the literature and are often referred to by different names across various fields of application. Although terminology may differ, the fundamental objective remains the same: to jointly characterize the susceptible and non-susceptible subpopulations within a unified modeling framework. Therefore, the presence of long-term survivors poses a limitation for conventional survival models, which do not explicitly account for a cured fraction in the population. However, a class of flexible survival models that accommodates both long- and short-term survivors has attracted increasing attention in the literature. Notably, these include the standard mixture model introduced by \cite{boag1949maximum} and subsequently extended by \cite{berkson1952survival}, as well as the promotion time model proposed by \cite{yakovlev1996}. The mixture cure rate model considered in this paper assumes that the population under study is heterogeneous, comprising both cured and non-cured individuals. Since its introduction, several extensions of the standard mixture cure rate model have been proposed in the literature; see, for example, \cite{de2022bayesian} and \cite{nakano2025mixture}.
	
	A common approach in survival analysis, whether or not a cure fraction is included, is to assume that all experimental units are susceptible to a single event of interest. However, in many practical scenarios, subjects may be at risk for multiple competing causes of failure, so that the occurrence of one event precludes or alters the probability of observing the others. Examples arise in several fields. In biomedical research, individuals may experience infection, disease relapse, or death from causes unrelated to the primary condition. In insurance and actuarial science, competing events may include accidents, theft, and natural disasters. In finance and economics, a company may exit the market because of bankruptcy, merger, or voluntary closure. In these settings, the occurrence of one event precludes the observation of alternative events as the first event of interest, thereby defining the competing risks framework.
	
	In the absence of competing risks, the Kaplan--Meier estimator \citep{kaplan1958nonparametric} is commonly used to assess the potential presence of a long-term survivor fraction. When multiple competing causes of failure are present, cumulative incidence functions offer a natural framework for characterizing the probabilities of different event types over time. A plateau below unity in the overall cumulative incidence may suggest that a proportion of the population remains free from all considered events in the long term. Nonetheless, this pattern alone is not sufficient to confirm the existence of a cured subpopulation, particularly when the follow-up period is limited.

	\subsection{Motivation}
	
	Cancer survival studies must account for substantial heterogeneity in patient outcomes. Patients with the same cancer diagnosis may experience different trajectories, such as death from the disease under study, death from other causes, or prolonged periods without any observed event. This heterogeneity may reflect underlying differences in sociodemographic and clinical characteristics. Models that focus solely on a single event may be inadequate when multiple causes of failure compete, and a proportion of the population may remain event-free in the long term. These considerations motivate the use of survival models that jointly address both competing risks and the presence of a cure fraction. By distinguishing the probability of susceptibility to each cause from the time to event among susceptible individuals, such models provide a comprehensive framework for characterizing cause-specific outcomes and long-term survival in population-based studies. The availability of large, population-based cancer databases provides an opportunity to investigate survival patterns in real-world populations. In particular, the “{\it Incidence SEER 2003–2023}” database, maintained by the National Cancer Institute of the United States of America, includes sociodemographic and clinical information on patients diagnosed with malignant skin cancer, along with data on vital status and cause of death. This resource supports the investigation of long-term survival patterns and the impact of competing causes of death among cancer patients.
	
	In SEER-based research, causes of failure are typically classified as death from melanoma or death from other causes. The latter category includes all deaths not attributed to melanoma, such as those resulting from other neoplasms, cardiovascular diseases (including myocardial infarction, heart failure, and stroke), chronic respiratory diseases, infections, metabolic diseases, pneumonia, and external causes like accidents, suicides, homicides, and any other unspecified reasons. Patients who do not experience failure from either cause during the follow-up period are considered right-censored observations. Among skin cancers, melanoma arises from melanocytes, the cells responsible for melanin synthesis. Although less common than non-melanoma skin cancers, melanoma is notable for its higher metastatic potential and, consequently, a substantially greater risk of mortality \citep{arnold2022global}.

	Prognosis in melanoma is strongly associated with disease stage at diagnosis. Patients diagnosed at earlier stages generally have higher survival probabilities, whereas those diagnosed at more advanced stages face greater risks of disease progression and mortality \citep{gershenwald2017melanoma}. Monitoring incidence and prevalence trends is therefore important for characterizing changes in disease occurrence and the number of individuals living with melanoma. These indicators can inform assessments of healthcare needs and the planning of long-term treatment and follow-up strategies. Early diagnosis remains important for improving outcomes, while identifying patients with advanced-stage disease is essential for appropriate clinical management.
	
	\vspace{-0.5cm}
	\begin{figure}[H]
		\centering
		$$
		\begin{array}{c}
			\includegraphics[scale=0.35]{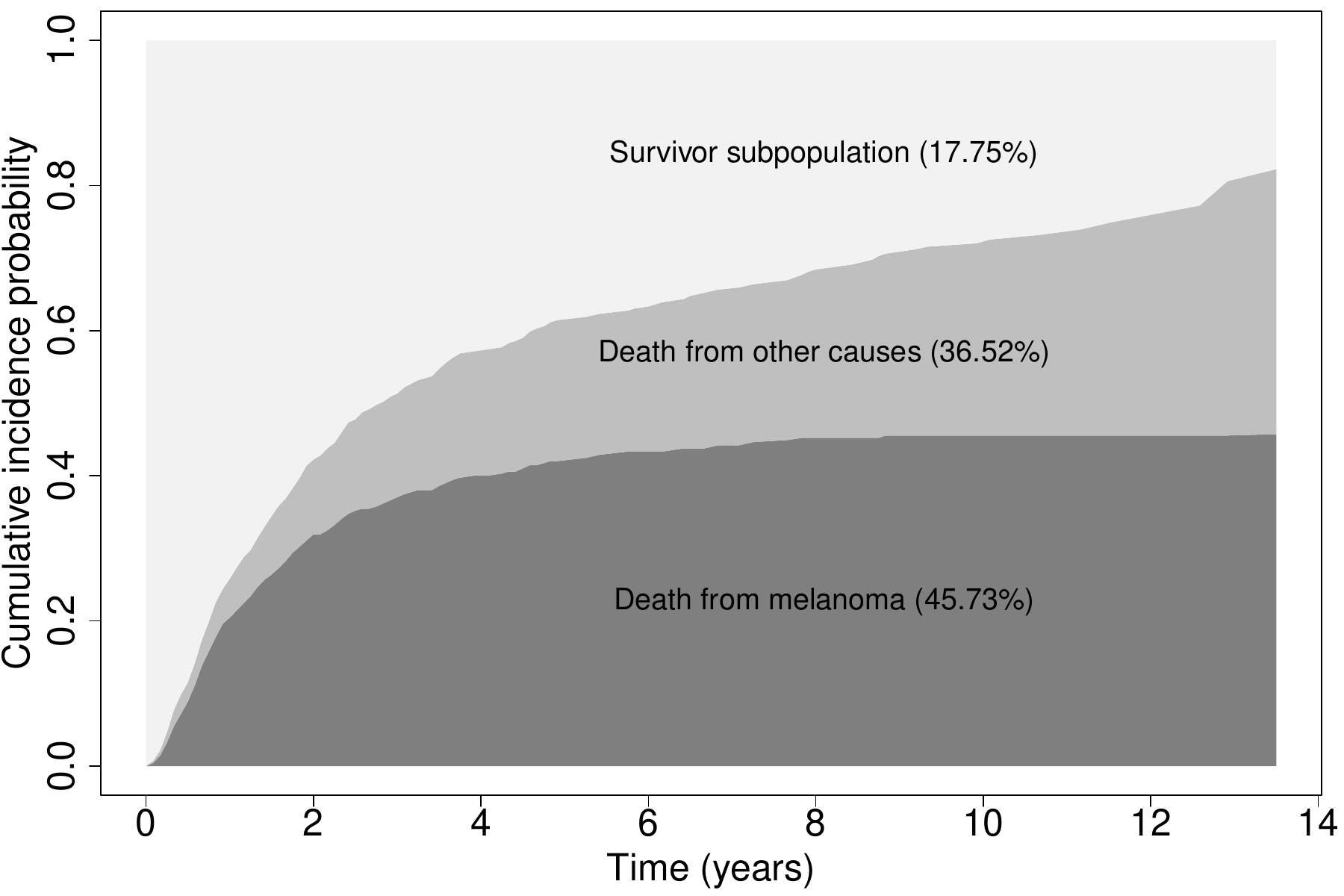}\\
		\end{array}
		$$
		\vspace{-0.8cm}
		\caption{Estimated cumulative probability incidence.}
		\label{cifAll}
	\end{figure}
	
	Figure~\ref{cifAll} presents the estimated cumulative incidence functions for the two competing causes of death in the SEER dataset. During the observed follow-up period, 45.7\% of individuals died from melanoma, 36.5\% died from other causes, and 17.8\% were censored. Thus, the total observed proportion of deaths was 82.2\%, with 17.8\% of individuals remaining event-free by the end of follow-up. While this distribution is consistent with the presence of a long-term survivor subpopulation, the observed censoring proportion alone does not confirm the existence or magnitude of a cure fraction. Nonetheless, the presence of competing causes of death and the potential for long-term survival motivate the use of survival models that jointly account for cure fractions and competing risks. 
	
	Previous studies have investigated the joint modeling of cure fractions and competing risks using parametric and semiparametric approaches \citep{ng2003based,gershenwald2017melanoma,choi2018semiparametric,ganguly2026model}. Nevertheless, further methodological advancements are necessary to address heterogeneous risk patterns among susceptible individuals. The specification of the latent event-time distribution is particularly critical, as it directly influences the hazard function and the range of survival patterns that the model can represent. To address these challenges, we propose a finite mixture parametric cure rate model with competing risks that incorporates flexible specifications for the latent event-time distribution. This framework enables a systematic exploration of how different distributional assumptions affect the characterization of survival outcomes in the presence of competing risks.

	In this article, we propose an approach to address the aforementioned statistical challenges to model the survival data in the presence of a cure fraction and multiple competing causes of the event of interest. The main objectives of this study are fourfold: $(i)$ to formulate a parametric cure rate model that accommodates competing risks; $(ii)$ to develop an estimation procedure utilizing the expectation-maximization (EM) algorithm; $(iii)$ to assess the finite-sample performance of the proposed estimators through an extensive Monte Carlo simulation study; and $(iv)$ to illustrate the practical applicability of the proposed methodology using real-world data involving a potential cure fraction and competing risks.
	
	The remainder of this paper is structured as follows. Section~\ref{sec2} introduces the proposed parametric cure rate model with competing risks and describes the estimation procedure. Section~\ref{sec3} evaluates the finite-sample performance of the proposed methodology through an extensive Monte Carlo simulation study. Section~\ref{sec4} illustrates its practical applicability using a real-world melanoma dataset. Finally, Section~\ref{sec5} presents the concluding remarks.

	\section{Competing risks-cure rate model}\label{sec2}
	
	Suppose that, during the follow-up period, each experimental unit may experience one of $J$ distinct and mutually exclusive failure types. Let $T_j>0$ be an absolutely continuous random variable denoting the latent time to failure from cause $j$, for $j = 1, \ldots, J$. Under the standard competing risks framework, the observed event time is $T^{*} = \min(T_1, \ldots, T_J)$, and the observed cause corresponds to the cause associated with the minimum latent event time. For the susceptible subpopulation, the cause-specific hazard function for cause $j$ is defined as
	
	\begin{equation}
		\lambda_j(t)= \displaystyle\lim_{\Delta t \to 0} \frac{\mathbb{P}(t\leq T^{*} < t + \Delta t, J=j|T^{*}\geq t)}{\Delta t}.
		\label{cspecifi.has}
	\end{equation}
	
	\noindent Based on expression~\eqref{cspecifi.has}, the cumulative incidence (or sub-distribution) function and the corresponding cause-specific survival function are defined, respectively, as $F_j\left(t\right) = \mathbb{P}\left(T^{*} \leq t, J = j\right)$ and $S_j\left(t\right) = \exp\left(-\int_{0}^t \lambda_j(y)\,dy\right)$. Additionally, the following relationship holds: $S\left(t\right) = \exp\left(-\sum\limits_{j=1}^J \int\limits_{0}^t \lambda_j(y)\,dy\right)$ and $\sum\limits_{j=1}^{J} F_j\left(t\right) + S\left(t\right) = 1$. Likewise, the cause-specific probability density function is given by $f_j\left(t\right) = \lambda_j\left(t\right) S\left(t\right)$, where $S(t)$ is the overall survival function up to time $t$.

	\subsection{Model formulation} \label{subsec2.1}
	
	Let $T^{*}$ denote the survival time for individuals susceptible to the event, modeled as a nonnegative, absolutely continuous random variable satisfying $T^{*}<\infty$ almost surely. For non-susceptible individuals, let $G_0=\infty$ with probability one. Define $Z$ as the susceptibility indicator variable, where $Z=1$ for uncured individuals and $Z=0$ for cured individuals. The combined survival time is then represented as $T = ZT^{*} + (1-Z) G_0$. Under the standard mixture cure rate model assumption, the overall survival function is given by:
	
	\begin{equation}
		S_{pop}\left(t|\bm{\psi}\right)=1-\pi\left(\bm{\xi}\right)+\pi\left(\bm{\xi}\right)S\left(t|\bm{\theta}\right),
		\label{simpleSMM}
	\end{equation}
	\vspace{0.05cm}

	\noindent where $S\left(t|\bm{\theta}\right)$ denotes the survival function of susceptible subjects, and $\pi\left(\bm{\xi}\right)=\mathbb{P}\left(Z=1|\bm{\xi}\right)$ is the susceptibility proportion in the mixture subpopulation. In addition, $\bm{\psi}=\left(\bm{\xi}^\top,\bm{\theta}^\top\right)^\top$ is the $d$-dimensional parameter vector, with $\bm{\xi}$ and $\bm{\theta}$ denoting the $(q+1)$- and $(p+1)$-dimensional parameter vectors affecting the cure and lifetime distributions, respectively. Although logistic regression is frequently chosen to model the susceptibility probability, $\pi\left(\bm{\xi}\right)$, due to the straightforward interpretation of its coefficients as odds ratios, other link functions, such as the probit and complementary log-log, are also viable options. Importantly, Equation~\eqref{simpleSMM} demonstrates that $S_{pop}\left(t|\bm{\psi}\right)$ is an improper survival function, since $\lim\limits_{t\rightarrow \infty} S_{pop}\left(t|\bm{\psi}\right) = 1 - \pi\left(\bm{\xi}\right) \neq 0$; thus, it does not satisfy the requirement of converging to zero for a proper survival function.
	
	In the competing risks setting, we extend the previously introduced framework by partitioning the susceptible population into $J$ mutually exclusive subgroups, each corresponding to a distinct cause of failure. Thus, the overall population consists of $(J+1)$ subgroups: one cured subgroup and $J$ cause-specific susceptible subgroups. Let $J_i \in \{0, 1, \ldots, J\}$ denote the subgroup membership indicator for individual $i\in \{1,2,\cdots,n\}$, where $J_i = 0$ indicates that individual $i$ is cured, and $J_i = j$ indicates susceptibility to cause $j$, for $j = \{1,2,\ldots, J\}$. The overall survival time for individual $i$ can then be expressed as $T_i = \sum\limits_{j=1}^{J} \mathbb{I}(J_i = j)T_{ij}^{\dagger} + \mathbb{I}(J_i=0)G_{0i}$, where $T_{ij}^{\dagger}$ denotes the latent failure time associated with cause $j$, and $G_{0i} = \infty$ almost surely for cured individuals. Under this specification, the population survival function in the finite-mixture cure rate model with competing risks can be expressed as:

	\begin{equation}
		S_{pop}\left(t_i|\bm{\psi}\right)=\pi_0\left(\bm{\xi}\right)+\sum\limits_{j=1}^J\pi_j\left(\bm{\xi}\right)S_j\left(t_i|\bm{\theta}_j\right),
		\label{SMM_CompRisks}
	\end{equation}
	
	\noindent 
	where $\pi_0(\bm{\xi}) = 1-\sum\limits_{j=1}^J \pi_j(\bm{\xi})$ denotes the cured proportion, and $S_j(t_i|\bm{\theta}_j) = \mathbb{P}(T_i > t_i, J_i = j|\bm{\theta}_j)$ represents the probability of surviving past time $t_i$ without experiencing cause $j$. Now, $\bm{\xi} = \left(\bm{\xi}_{1}^\top, \bm{\xi}_{2}^\top, \ldots, \bm{\xi}_{J}^\top\right)^\top$ and $\bm{\theta} = \left(\bm{\theta}_{1}^\top, \bm{\theta}_{2}^\top, \ldots, \bm{\theta}_{J}^\top\right)^\top$ are $(q+1)\times J$ and $(p+2)\times J$-dimensional parameter vectors governing the cure and lifetime distributions, respectively. Finally, $\bm{\psi} = \left(\bm{\xi}^{\top}, \bm{\theta}^{\top}\right)^\top$ is the $d$-dimensional model parameter vector, with $d = (q+p+3)\times J$. For this setting, the incidence component is modeled through multinomial logistic regression, whereby $\pi_j(\bm{\xi}) = \mathbb{P}\left(T_i<\infty, J_i = j|\bm{\xi}\right) = \exp\left(\mathbf{w}_i^\top \bm{\xi}_j\right)/\sum\limits_{k=0}^{J} \exp\left(\mathbf{w}_i^\top \bm{\xi}_k\right)$, where $\bm{\xi}_j = \left(\xi_{0j}, \xi_{1j}, \cdots, \xi_{qj}\right)^\top$ and $\mathbf{w}_i = (w_{i0}, w_{i1}, \cdots, w_{iq})^\top$ is the vector of covariates affecting the cure distribution, with $w_{i0}=1$ for all $i\in \{1,2,\cdots,n\}$. To ensure model identifiability, we set $\bm{\xi}_0 = \mathbf{0}_{q+1}$. Furthermore, the density and hazard functions for the overall population are defined as $f_{pop}\left(t_i|\bm{\psi}\right) = -\frac{\partial}{\partial t_i} S_{pop}\left(t_i|\bm{\psi}\right)$ and $\lambda_{pop}\left(t_i|\bm{\psi}\right) = -\frac{\partial}{\partial t_i} \log S_{pop}\left(t_i|\bm{\psi}\right)$, respectively.
	
	Equation~\eqref{SMM_CompRisks} represents the population survival function as a mixture of the cured subgroup and $J$ susceptible subgroups, each corresponding to a distinct cause of failure. In the special case where $J = 1$, this formulation reduces to the single-cause model presented in Equation~\eqref{simpleSMM}. Covariates are included in both parts of the model to reflect their possible impact on an individual’s susceptibility to the $j$th event, as well as on the timing of event occurrence among those who are susceptible.
	
	The regression structure is incorporated into the cause-specific survival function $S_j\left(t_i|\bm{\theta}_j\right)$ as follows. Let $\mathbf{x}_i = \left(x_{i0}, x_{i1}, \ldots, x_{ip}\right)^\top$ denote the covariate vector for individual $i$, where $x_{i0} = 1$ represents the intercept. The corresponding regression coefficient vector is $\bm{\beta}_j=\left(\beta_{0j}, \beta_{1j}, \ldots, \beta_{pj}\right)^\top \in \mathbb{R}^{p+1}$, which forms part of the parameter vector $\bm{\theta}_j$, for $j = 1, 2, \ldots, J$.
	Under the accelerated failure time framework, the logarithm of the latent failure time for an individual susceptible to cause $j$ is modeled as $\log T_{\mathbf{x}_i, j}^{*} = \mu_j(\mathbf{x}_i) + \sigma_j W_j$, where $\mu_j(\mathbf{x}_i) = \mathbf{x}_i^\top \bm{\beta}_j$ is the location parameter, $\sigma_j > 0$ is the scale parameter, and $W_j$ is a random error term. Thus, the log-survival time follows a location-scale model. To accommodate different distributional forms, we consider three common specifications for $W_j$: the Gumbel, Logistic, and Log-gamma distributions.

	\subsection{Parameters estimation}\label{subsec2.2}
	
	In this section, we derive the maximum likelihood estimator of $\bm{\psi}$. Let $D_{oi}=\{\left(t_i,\delta_i,J_i,\mathbf{x}_i, \mathbf{w}_i\right),\,\,i=1,2,\cdots,n\}$ denote the observed dataset. For subject $i$, $t_i$ denotes the observed survival time for the overall population, and $\delta_i$ indicates whether a failure event was observed. The vectors $\mathbf{x}_i$ and $\mathbf{w}_i$, of dimensions $(q+1)$ and $(p+1)$ respectively, contain the covariates used in the survival and cure components of the model. The cause-specific event indicator is defined as $\delta_{ij}=\delta_i \mathbb{I}\left(J_i=j\right),\,\, j = 1, 2, \ldots, J$.  Let $T_{i}^{*}=\min\limits_{1 \leq j \leq J} \{T_{ij}^{\dagger}\}$ denote the latent failure time for subject $i$ associated with cause $j$, and let $C_i$ denote the censoring time. The observed time is $t_i = \min\left(T_{i}^{*}; C_i\right)$ for susceptible individuals, and $t_i=G_{0i}$ otherwise. In addition, the failure indicator is defined as $\delta_i=\mathbb{I}\left(T_{i}^{*} \leq C_i\right)$ for susceptible subjects and $\delta_i=0$ otherwise. Under the non-informative censoring mechanism, the observed-data log-likelihood is given by
	
	\vspace{-0.4cm}
	{\small
		\begin{equation}
			\ell(\bm{\psi}|D_{o})=\sum\limits_{i=1}^n\left\lbrace\sum\limits_{j=1}^J\delta_{ij}\log\left[\pi_j\left(\bm{\xi}\right) \, f_j(t_i|\bm{\theta}_j)\right]+\left(1-\delta_{i}\right)\log\left[\sum\limits_{j=0}^J \pi_j\left(\bm{\xi}\right)\, S_j\left(t_i|\bm{\theta}_j\right)\right]\right\rbrace.
			\label{obsLik}
		\end{equation}
	}
	
	\noindent The MLE of $\bm{\psi}$ is obtained by maximizing the log-likelihood function given in Equation~\eqref{obsLik}, i.e., $\hat{\bm{\psi}}=\operatorname*{argmax}\limits_{\bm{\psi} \in \mathbb{R}^{d}}\,\ell(\bm{\psi}|D_{o})$. However, directly maximizing the observed-data likelihood can be challenging because of the model’s mixture structure, especially when many competing risks are present. To facilitate estimation, we use the complete-data likelihood, whose factorized form provides a convenient foundation for implementing EM-type algorithms~\citep{dempster1977maximum} and for separately estimating the different model components.
	
	The rationale behind the EM algorithm is straightforward. Formally, let $D_{c} = (D_{o}, \mathbf{Z})$ denote the complete data, where $\mathbf{Z} = (Z_{ij})$ is an $n \times J$ matrix of latent indicators, with $\mathbf{Z}_j=(Z_{1j}, Z_{2j}, \ldots, Z_{nj})^\top$ and $Z_{ij} = \mathbb{I}\left(J_i=j\right)$. For individuals with $\delta_{ij} = 0$, $Z_{ij}$ is a latent variable indicating cure status ($Z_{ij} = 0$) or susceptibility to failure from cause $j$ ($Z_{ij} = 1$). Under the augmented data notation and the cause-specific competing risks framework, the overall population probability density and survival functions can be written as follows, $f_{\mathrm{pop}}(t_i|\bm{\psi}) = \prod\limits_{j=1}^{J}\left[\pi_j(\bm{\xi}) f_j(t_i|\bm{\theta}j)\right]^{Z{ij}}$ and $S_{\mathrm{pop}}(t_i|\bm{\psi}) = \pi_0(\bm{\xi})^{Z_{i0}} \prod\limits_{j=1}^{J} \left[ \pi_j(\bm{\xi}) S_j(t_i|\bm{\theta}_j)\right]^{Z{ij}}$, respectively. Under the non-informative censoring scheme, the complete-data likelihood function is then given by:
	
	\vspace{-0.3cm}
	{\small
		\begin{align}\label{completeLik} 
			L_c(\bm{\psi}|D_c) &\propto \prod\limits_{i=1}^n \left[\prod\limits_{j=1}^J\left\lbrace\pi_j\left(\bm{\xi}\right)f_j\left(t_i|\bm{\theta}_j\right)\right\rbrace^{Z_{ij}}\right]^{\sum\limits_{j=1}^J\delta_{ij}}  \left[\pi_0\left(\bm{\xi}\right)^{Z_{i0}}\prod\limits_{j=1}^J\left[\pi_j\left(\bm{\xi}\right)S_j\left(t_i|\bm{\theta}_j\right)\right]^{Z_{ij}}\right]^{1-\sum\limits_{j=1}^J\delta_{ij}} \nonumber\\
			&=\prod\limits\limits_{i=1}^n \left[\prod\limits_{j=1}^J\pi_0\left(\bm{\xi}\right)^{Z_{i0}}\pi_j\left(\bm{\xi}\right)^{Z_{ij}}\prod\limits_{j=1}^J\lambda_j\left(t_i|\bm{\theta}_j\right)^{\delta_{ij}}S_j\left(t_i|\bm{\theta}_j\right)^{Z_{ij}}\right] \nonumber\\
			&=\underset{L_{Inc}(\boldsymbol{\xi}|D_c)}{\underbrace{\prod\limits_{i=1}^n \prod\limits_{j=1}^J\pi_0\left(\bm{\xi}\right)^{Z_{i0}}\pi_j\left(\bm{\xi}\right)^{Z_{ij}}}}\underset{L_{Lat}(\bm{\theta}|D_c)}{\underbrace{    \prod\limits_{i=1}^n\,\,\prod\limits_{j=1}^J \lambda_j\left(t_i|\bm{\theta}_j\right)^{\delta_{ij}}S_j\left(t_i|\bm{\theta}_j\right)^{Z_{ij}}}}.
		\end{align}
	}
	
	\noindent From Equation~\eqref{completeLik}, the complete likelihood function is given by:
	$\ell_c\left(\bm{\psi}|D_c\right) = \ell_{\text{Inc}}\left(\bm{\xi}|D_c\right) + \ell_{\text{Lat}}\left(\bm{\theta}|D_c\right)$, where $\ell_{\text{Inc}}\left(\bm{\xi}|D_c\right)=\sum\limits_{i=1}^n\sum\limits_{j=0}^JZ_{ij}\log\left(\pi_j\left(\bm{\xi}\right)\right)$, with $Z_{i0}=1-\sum\limits_{j=1}^{J}Z_{ij}$ and $\ell_{\text{Lat}}\left(\bm{\theta}|D_c\right)=\sum\limits_{i=1}^n\sum\limits_{j=1}^J\delta_{ij}\log \lambda_{j}\left(t_i|\bm{\theta}_j\right)+\sum\limits_{i=1}^n\sum\limits_{j=1}^JZ_{ij}\log S_{j}\left(t_i|\bm{\theta}_j\right)$. Since the latent variable $Z_{ij}$ appears within a summation in both log-likelihood expressions, the expected value of the complete log-likelihood is $\tilde{\ell}_c\left(\bm{\psi}|D_o, \bm{\psi}^{(g)}\right) = \mathbb{E}\left[\ell_c\left(\bm{\psi}|D_c\right)|D_o, \bm{\psi}^{(g)}\right]$ and can be expressed as the sum of the following terms:
	
	\vspace{-0.4cm}
	\begin{eqnarray}
		\tilde{\ell}_{\text{Inc}}\left(\bm{\xi}|\mathcal{D}_{o},\bm{\tau}^{(g)}\right)&=&\sum\limits_{i=1}^n\sum\limits_{j=1}^J\tau_{ij}^{(g)}{\bf w}_i^\top\bm{\xi}_j-\sum\limits_{i=1}^n\log\left(\sum\limits_{k=0}^J\exp\left({\bf w}_i^\top\bm{\xi}_k\right)\right), \label{ExpectedLik.Inc}
		\\
		\tilde{\ell}_{\text{Lat}}\left(\bm{\theta}|\mathcal{D}_{o},\bm{\tau}^{(g)}\right)&=&\sum\limits_{i=1}^n\sum\limits_{j=1}^J\delta_{ij}\log \lambda_{j}\left(t_i|\bm{\theta}_j\right)+\sum\limits_{i=1}^n\sum\limits_{j=1}^J\tau_{ij}^{(g)}\log S_{j}\left(t_i|\bm{\theta}_j\right), \label{ExpectedLik.Lat}
	\end{eqnarray}

	\noindent denoting the expected log-likelihood functions for the cure and lifetime distributions, respectively. In the E-step, $\bm{\tau}^{(g)} = \left(\tau_{i1}^{(g)}, \tau_{i2}^{(g)}, \ldots, \tau_{iJ}^{(g)}\right)$ denotes the vector of posterior weights for a fixed $i$, where the preliminary weights $\tau_{ij}^{(g)}$ are first computed as 
	
	\vspace{-0.2cm}
	\begin{equation}
		\tau_{ij}^{*(g)}=\mathbb{E}\left(Z_{ij}=1|D_{o},\bm{\psi}^{(g)},\delta_{ij}=0\right)=\frac{\pi_j^{(g)}\left(\bm{\xi}\right)S_{j}\left(t_i|\bm{\theta}_j^{(g)}\right)}{\sum\limits_{k=0}^J\pi_k^{(g)}\left(\bm{\xi}\right)S_k\left(t_i|\bm{\theta}_k^{(g)}\right)},
		\label{EM.weight}
	\end{equation}
	
	\noindent and subsequently updated as $\tau_{ij}^{(g)} = \delta_{ij} + (1-\delta_{ij})\tau_{ij}^{(g)}$. Thus, $\tau_{ij}^{(g)}=1$ when the event is observed to be due to cause $j$, and $\tau_{ij}^{(g)}=\tau_{ij}^{*(g)}$ for censored observations. Additionally, $\tau_{i0}^{(g)} = 1 - \sum\limits_{j=1}^{J} \tau_{ij}^{(g)}$. The EM algorithm proceeds by alternating between the E-step and M-step until convergence. In the E-step, posterior weights are obtained using Equation~\eqref{EM.weight} and incorporated into the expected log-likelihood functions defined in Equations~\eqref{ExpectedLik.Inc} and~\eqref{ExpectedLik.Lat}. In the subsequent M-step, these functions are maximized to update the parameter estimates. The maximization can be performed separately for the incidence and latency components using numerical optimization methods such as the Newton–Raphson algorithm or the \texttt{optim} function in \texttt{R} \citep{RCoreTeam2026}. For the incidence component, multinomial logistic regression is fitted using the \texttt{multinom} function from the \texttt{nnet} package. The E- and M-steps are alternated until the prespecified convergence criterion is met.
	
	\subsection{Standard error estimation}\label{subsec2.3}
	
	Although the EM algorithm provides a convenient framework for obtaining the maximum likelihood estimator of $\bm{\psi}$, it does not directly yield the observed information matrix, $\mathcal{H}\left(\hat{\bm{\psi}}\right)$, which is required for variance estimation. A well-established solution to this problem is Louis’ method \citep{louis1982}, which expresses the observed information in terms of the conditional moments of the complete-data score and information, given the observed data. However, applying this method requires characterizing the conditional distribution of the latent variables given $D_o$, which can be analytically challenging in mixture cure models with competing risks due to their complex latent structure. To circumvent this complexity, we estimate the asymptotic variance-covariance matrix of $\hat{\bm{\psi}}$ by numerically evaluating the Hessian of the observed log-likelihood in Equation~\eqref{obsLik} at the maximum likelihood estimate. Specifically, the matrix
	
	\begin{equation}
		\Sigma_A^{-1}\left(\hat{\bm{\psi}}\right)=\left[-\frac{\partial^2\ell(\bm{\psi}|D_{o})}{\partial \bm{\psi}\partial \bm{\psi}^{\top}}\Bigg|_{\bm{\psi}=\hat{\bm{\psi}}}\right]^{-1}.
		\label{ObsInfo}
	\end{equation}
	
	\noindent can be evaluated numerically using the \texttt{hessian} function from the \texttt{pracma} package in {\texttt R}. Additionally, bootstrap standard errors were employed as a complementary measure of uncertainty and compared with the asymptotic standard errors derived from the inverse of the observed information matrix.
	
	The bootstrap procedure is particularly useful because it accounts for the variability arising from the entire estimation process, including the estimation of latent class probabilities and the parameters governing both the incidence and latency components \citep{efron1986,davison1997}. Accordingly, the bootstrap variance–covariance matrix of the parameter estimates is obtained jointly as
	
	\vspace{-0.3cm}
	\begin{equation}
		\Sigma_{\text{B}}^{-1}\left(\hat{\bm{\psi}}\right)=\frac{1}{B-1}\sum\limits_{b=1}^B \left(\hat{\bm{\psi}}^{(b)}-\bar{\hat{\bm{\psi}}}\right)\left(\hat{\bm{\psi}}^{(b)}-\bar{\hat{\bm{\psi}}}\right)^\top,
		\label{bootSE}
	\end{equation}
	
	\noindent where $\hat{\bm{\psi}}^{(b)}$ is the EM-based maximum likelihood estimate from the $b$-th bootstrap estimates, and $\bar{\hat{\bm{\psi}}} = \frac{1}{B} \sum\limits_{b=1}^{B} \hat{\bm{\psi}}^{(b)}$ is the mean across all $B$ bootstrap replications. Under standard regularity conditions, $\sqrt{n}\left(\hat{\bm{\psi}}-\bm{\psi}\right)\overset{\mathcal{D}}{\rightarrow}\mathcal{N}_d\left({\bf 0}_d, {\bf I}_F^{-1}\left(\bm{\psi}_0\right)\right)$ as $n\rightarrow \infty$, since both
	$n^{-1}\Sigma_{\text{A}}\left(\hat{\bm{\psi}}\right)$ and $n^{-1}\Sigma_{\text{B}}\left(\hat{\bm{\psi}}\right)$ converge in probability to ${\bf I}_F\left(\bm{\psi}_0\right)$, with $\bm{\psi}_0$ denoting the true parameter value of $\bm{\psi}$ and $\overset{\mathcal{D}}{\rightarrow}$ denotes the convergence in distribution.

	\section{Simulation study}\label{sec3}
	
	To investigate the finite-sample behavior of the proposed models, we conducted a series of Monte Carlo simulation studies. Each simulated datasets included two covariates in the regression structure: a binary predictor, $x_{1i} \sim \mathrm{Bernoulli}(0.5)$, and a continuous predictor, $x_{2i} \sim N(0,1)$. The corresponding covariate vector was specified as $\mathbf{x}_i = (1, x_{1i}, x_{2i})^\top$, for every $i\in \{1,2,\cdots,n\}$. To maintain a common regression structure across model components, the same covariates were used for both the incidence and latency submodels, so that $\mathbf{x}_i = \mathbf{w}_i$. Each scenario involved two competing causes of failure ($J = 2$), with the cause indicator $J_i$ drawn from a multinomial distribution with probabilities $\pi_j(\bm{\xi}) = \exp\left(\mathbf{w}_i^\top \bm{\xi}_j\right)/\sum\limits_{k=0}^{2} \exp\left(\mathbf{w}_i^\top \bm{\xi}_k\right)$, where $\pi_0(\bm{\xi}) = 1 - \pi_1(\bm{\xi}) - \pi_2(\bm{\xi})$ and $\bm{\xi}_j=\left(\xi_{0j},\xi_{1j},\xi_{2j}\right)^\top$. In this case, $J_i \in \{0, 1, 2\}$, where $J_i = 0$ indicates that the $i$-th subject is non-susceptible to both event types, while $J_i\neq 0$ indicates susceptibility to failure from cause 1 or cause 2, respectively.
	
	The latent survival times $T_{1i}^{\dagger}$ and $T_{2i}^{\dagger}$ were generated using the inverse transform method; specifically, $T_{ij}^{\dagger} = F_j^{-1}(u_i)$, where $j = 1, 2$ and $u_i \sim \mathcal{U}(0, 1)$. Here, $F_j(\cdot)$ denotes the cumulative incidence function for the $j$-th cause of failure. In the finite mixture cure model with competing risks, the observed time is defined as $t_i = \min\{T_i^{*}, C_i\}$ for susceptible individuals and $t_i = G_{0i}$ otherwise. We set $G_{0i} \equiv C_i$, where $C_i$ is the censoring time, generated from an exponential distribution with mean $\mathbb{E}\left(C_i\right) = \tau$. The term $T_i^{*} = \min\{T_{i1}^{\dagger}, T_{i2}^{\dagger}\}$ denotes the minimum of the two potential failure times. The failure indicator is defined as $\delta_i = w_i$ for susceptible individuals, where $w_i = \mathbb{I}\{T_i^{*} \leq C_i\}$ indicates whether a failure occurs before censoring, and $\delta_i = 0$ for non-susceptible individuals.
	
	To assess the finite-sample performance of the proposed model, four sample sizes were examined: $n \in {200, 600, 800, 1{,}000}$. Let $\psi_h\in \bm{\psi}$ denote the true value of the $h$-th model parameter, and let $\hat{\psi}_h^{(m)}$ be the corresponding maximum likelihood estimate in the $m$-th Monte Carlo sample, for $m = 1, 2, \ldots, M$. For each sample size, 1,000 Monte Carlo datasets were generated to evaluate the model performance. The assessment considered empirical bias (Bias), asymptotic standard error (ASE), bootstrap standard error (BSE), root mean squared error (RMSE), and empirical coverage probabilities calculated using asymptotic (ACP) and bootstrap (BCP) variance estimates, as described in Section~\ref{subsec2.3}. Under the asymptotic normality of \(\hat{\psi}_h\), the 95\% Wald-type confidence interval for $\psi_h$ are given by $
	\text{ACI}\left(\psi_h\right) = \hat{\psi}_h \pm 1.96 \left[\Sigma_{A}^{-1}\left(\hat{\bm{\psi}}\right)\right]_{hh}^{1/2}$
	and $\text{BCI}\left(\psi_h\right) = \hat{\psi}_h \pm 1.96 \left[\Sigma_{\text{B}}^{-1}\left(\hat{\bm{\psi}}\right)\right]_{hh}^{1/2}$. Here, $\left[\Sigma_{\text{A}}^{-1}\left(\hat{\bm{\psi}}\right)\right]_{hh}$ and $\left[\Sigma_{\text{B}}^{-1}\left(\hat{\bm{\psi}}\right)\right]_{hh}$ denote the $(h,h)$ entries of the asymptotic and bootstrap variance–covariance matrices, respectively. The BSE was computed using Equation~\eqref{bootSE}, based on 350 bootstrap resamples per Monte Carlo replication. All performance measures were evaluated over 1,000 Monte Carlo replications, and the coverage probability was estimated as the proportion of replications in which the Wald-type confidence interval contained the true parameter value, $\psi_h$.
	
	As previously discussed in Section~\eqref{subsec2.1}, for the statistical modeling of the cause-specific hazard function, three distributions were considered for the survival times $T_{ij}^{\dagger}$: $(i)$ $T_{ij}^{\dagger}\sim\mathrm{Weibull}\,(\alpha_j, \phi_{ij})$, $(ii)$ $T_{ij}^{\dagger}\sim\mathrm{Gamma}\,(\alpha_j, \phi_{ij})$, and $(iii)$ $T_{ij}^{\dagger} \sim \mathrm{Log\text{-}Logistic}\,(\alpha_j, \phi_{ij})$, where $\alpha_j>0$, $\phi_{ij} = \exp\left({\bf x}_i^\top \bm{\beta}_j\right)$ and $\bm{\beta}_j=\left(\beta_{0j}, \beta_{1j}, \beta_{2j}\right)^\top$. 
	Two simulation scenarios were considered, with cure fractions of approximately $\pi_0(\bm{\xi}) = 0.40$ (Scenario~1) and $\pi_0(\bm{\xi}) = 0.30$ (Scenario~2). To obtain an overall censoring rate of about 50\% in the regression models under both scenarios, we fixed $\tau = 5$.

	\begin{table}[H]
		\sf
		\footnotesize
		\setlength{\tabcolsep}{0.10cm}
		\centering
		\caption{Scenarios considered for the three parametric models.}
		\vspace{-0.2cm}
		\begin{tabular}{ccrrrrrrrrrrrrrrrrr}
			\hline
			\multicolumn{1}{c}{\multirow{2}{*}{\makecell{Regression\\ model}}} & \multicolumn{1}{c}{\multirow{2}{*}{Scenario}} &&\multicolumn{7}{c}{\rule[-1ex]{0pt}{3ex}  Cause 1}  && \multicolumn{7}{c}{Cause 2}\\ 
			\cline{4-10} \cline{13-19} 
			\multicolumn{1}{l}{}  & \multicolumn{1}{l}{}                         
			&& $\beta_{01}$ & $\beta_{11}$ & $\beta_{21}$ & $\alpha_{1}$ & $\xi_{01}$ & $\xi_{11}$ & $\xi_{21}$ &&& $\beta_{02}$ & $\beta_{12}$ & $\beta_{22}$ & $\alpha_{2}$ & $\xi_{02}$ & $\xi_{21}$ & $\xi_{22}$\\
			\hline
			\multirow{2}{*}{Weibull}
			& 1 && -1 & 0.8 &-1 & 0.8 &-1 & 0.8& 1&&& -1& 0.8 &-1 &1.2 & 1 & 0.8 & -1\\
			& 2 && -1 & 0.8 &-1 & 0.8 & 1 &-0.8&-1&&& -1& 0.8 & 1 &1.2 & 1 &-0.8 &  1\\
			\hline
			\multirow{2}{*}{Gamma}
			& 1 &&-1& 0.8& -1& 0.8 & -1& 0.8 & 1  &&& -1& 0.8&-1  &1.2 &-1& 0.8& -1 \\
			& 2 &&-1& 0.8& -1& 0.8 &-1& 0.8 & 1  &&&  1& -0.8&-1  &1.2 & 1&-0.8&  1\\
			\hline
			\multirow{2}{*}{Log-Logistic}
			& 1 && -1& -0.8& -1&0.8 &-1 & 0.8& 1 &&& -1&-0.8&-1  &1.2 & -1& 0.8& -1\\
			& 2 &&  1& -0.8& -1&1.2 &-1 & 0.8&-1 &&& -1& 0.8& 1  &0.8 &  1&-0.8& -1\\
			\hline
		\end{tabular}
		\label{scenarios}
	\end{table}
	
	\noindent The true parameter configurations for each scenario and across the two competing causes are reported in Table~\ref{scenarios}. These configurations yield distinct patterns in the overall hazard function, $\lambda\left(t_i|\bm{\theta}\right) = \lambda_1\left(t_i|\bm{\beta}_1, \alpha_1\right) + \lambda_2\left(t_i|\bm{\beta}_2, \alpha_2\right)$. Specifically, under the Weibull regression model, the hazard function decreases for the first cause and increases for the second cause in both scenarios, resulting in a classic bathtub-shaped hazard function. For the Gamma distribution, the cause-specific hazard function also decreases for cause 1 and increases for cause 2. The Log-logistic model displays distinct patterns across the two scenarios. The estimated proportions of cured individuals also vary by scenario and model. In Scenario 1, approximately 25\% of cured individuals belong to the control group ($x_{1i}=0$), compared with 15\% in the treatment group ($x_{1i}=1$); this pattern is consistent across all three models. In Scenario 2, the pattern reverses, with the treatment group accounting for approximately 20\% of cured individuals and the control group for 10\%.
	
\begin{figure}[H]
		\centering
		\includegraphics[scale=0.39]{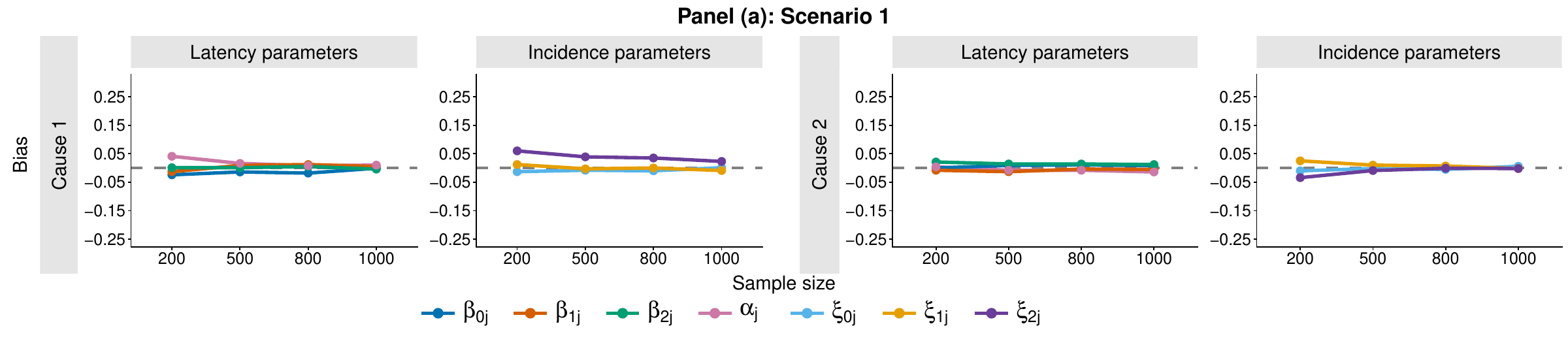}\\
		\includegraphics[scale=0.39]{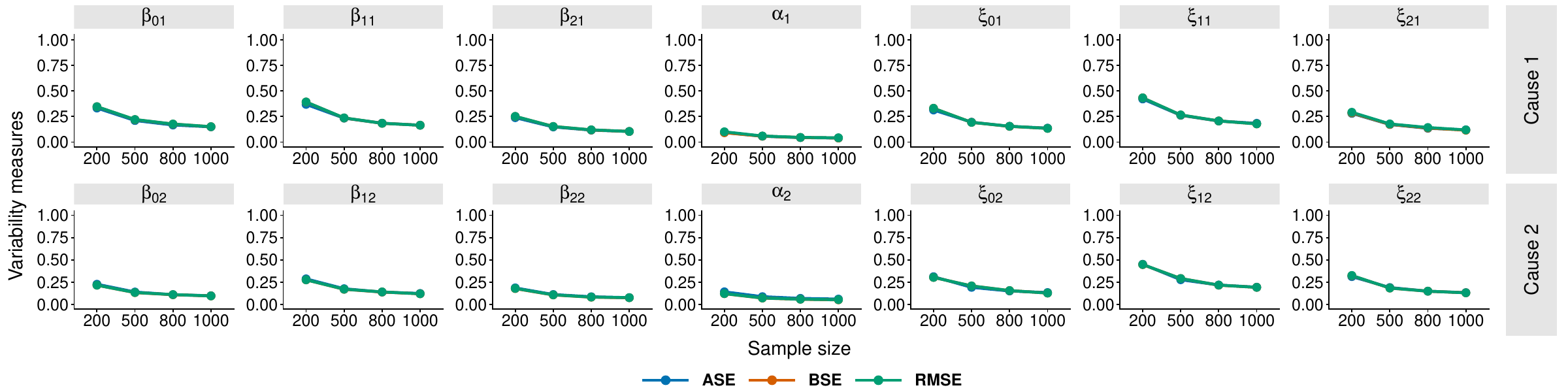}\\
		\includegraphics[scale=0.39]{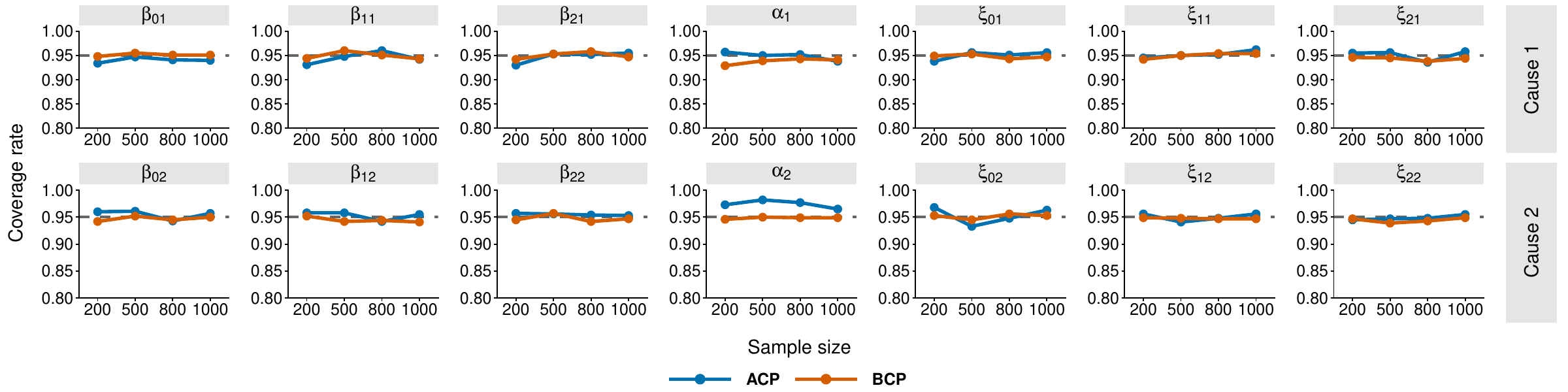}\\
		\hspace{0.01cm}
		\includegraphics[scale=0.39]{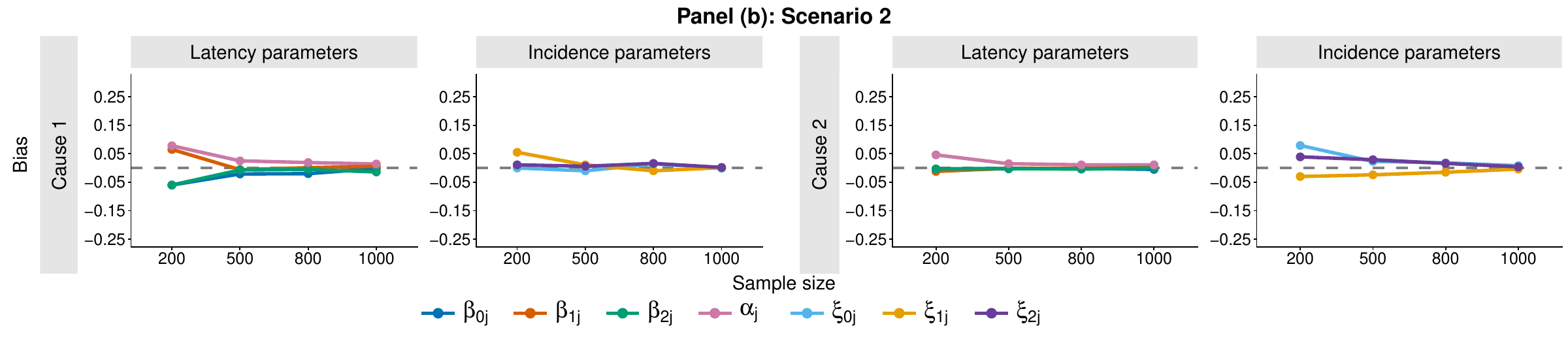}\\
		\includegraphics[scale=0.39]{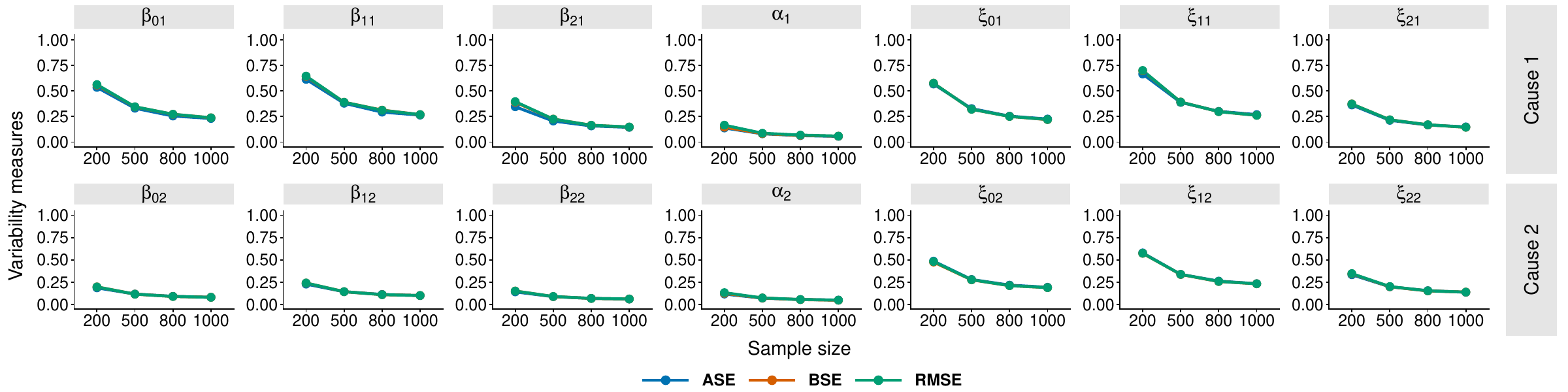}\\
		\includegraphics[scale=0.39]{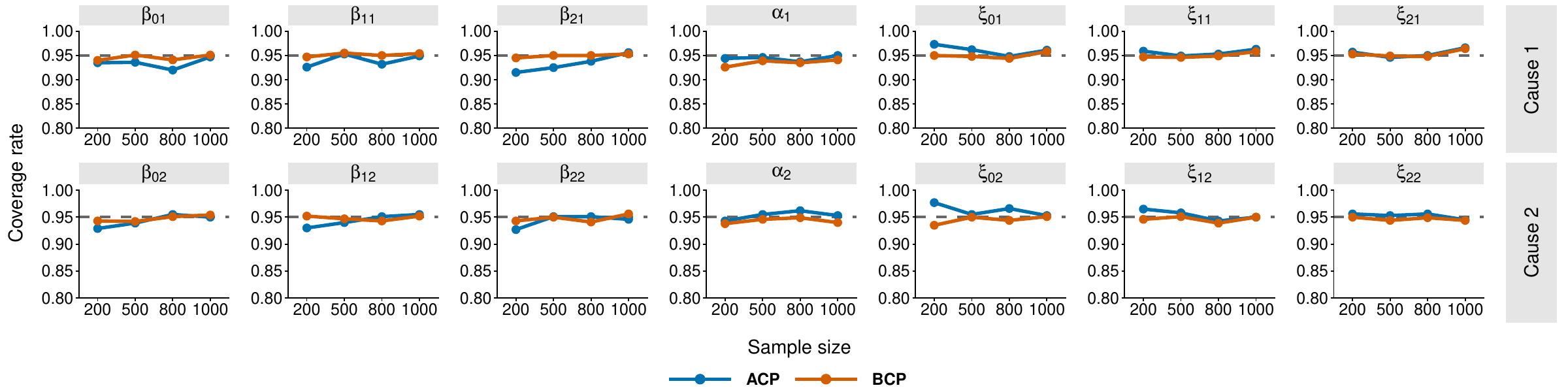}\\
		\vspace{-0.2cm}
		\caption{Monte Carlo simulation results for the Weibull-Multinomial mixture cure rate regression model for both scenarios, for $j=1$ (cause 1) and $j=2$ (cause 2).}
		\label{SimuWB}
	\end{figure}

\begin{figure}[H]
		\centering
		\includegraphics[scale=0.39]{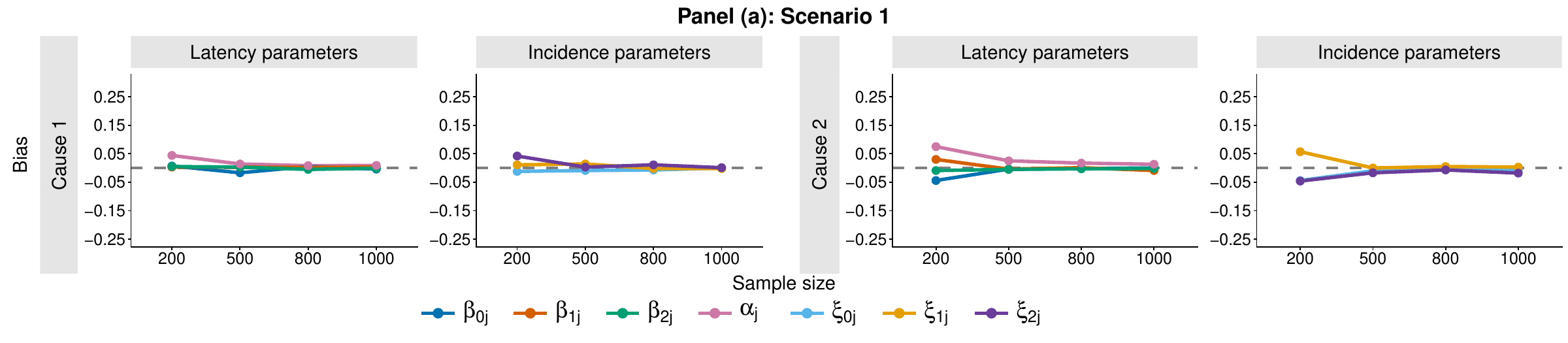}\\
		\includegraphics[scale=0.39]{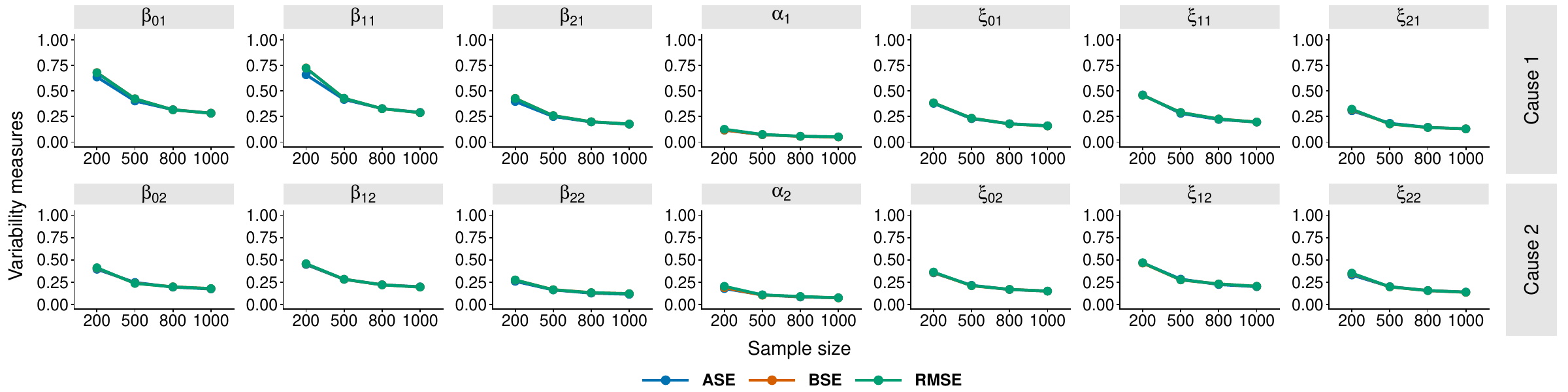}\\
		\includegraphics[scale=0.39]{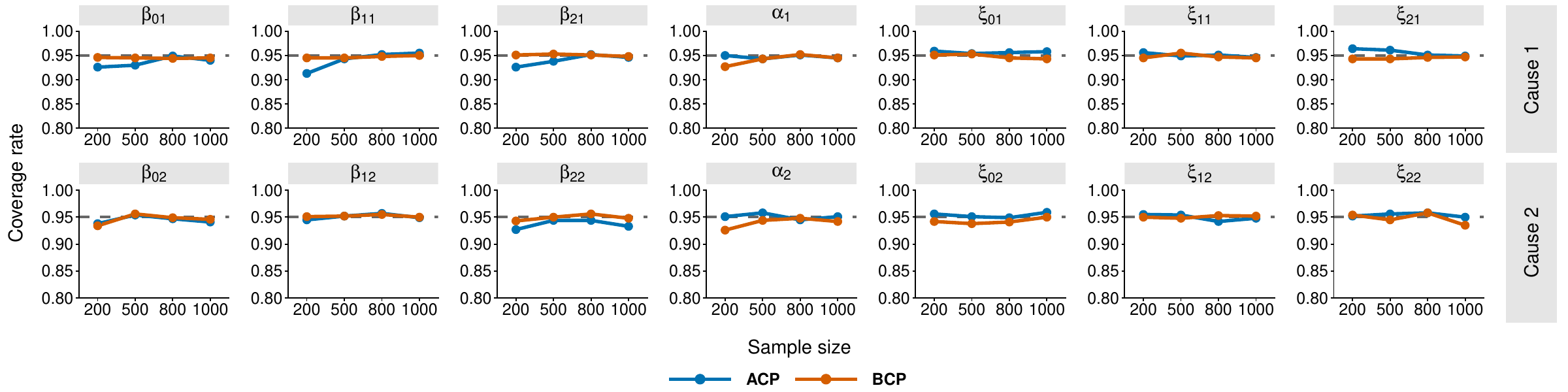}\\
		\hspace{0.01cm}
		\includegraphics[scale=0.39]{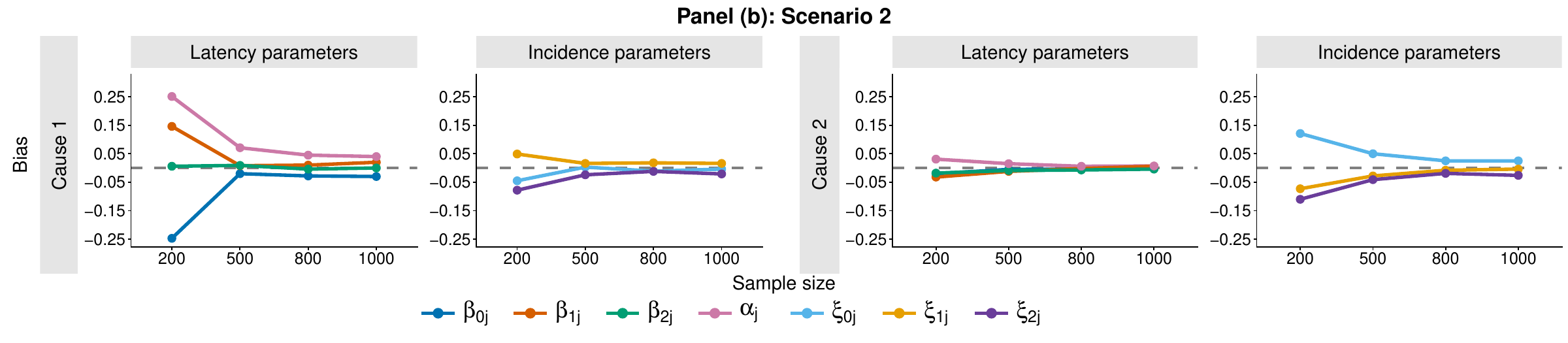}\\
		\includegraphics[scale=0.39]{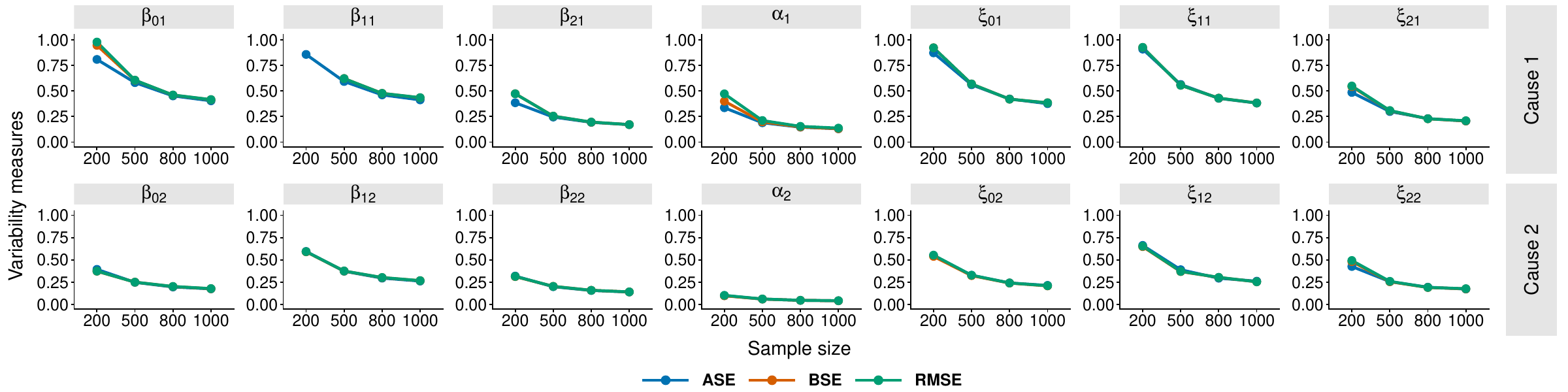}\\
		\includegraphics[scale=0.39]{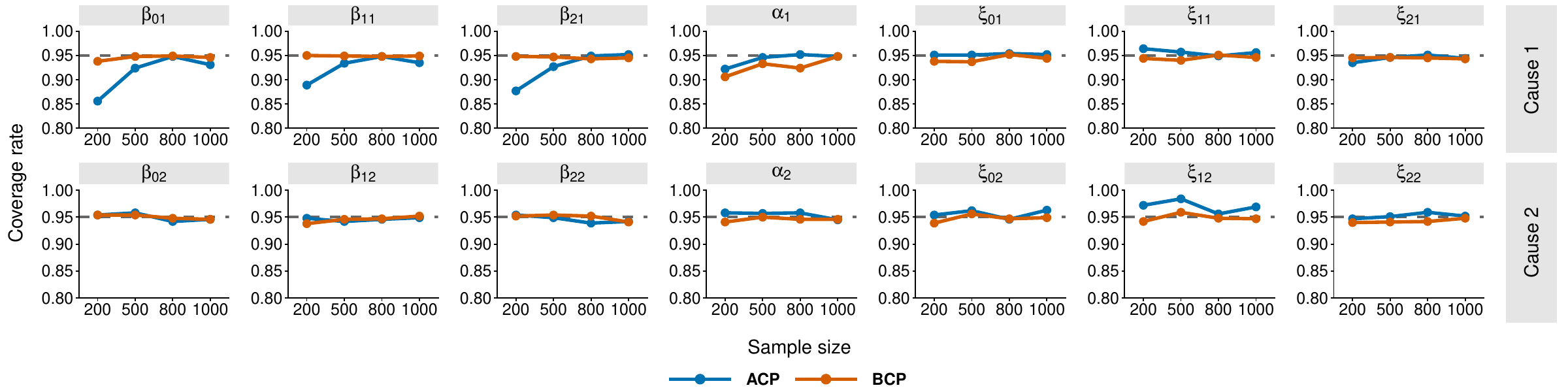}\\
		\vspace{-0.2cm}
		\caption{Monte Carlo simulation results for the Log-Logistic-Multinomial mixture cure rate regression model for both scenarios, for $j=1$ (cause 1) and $j=2$ (cause 2).
		}
		\label{SimuLogL}
	\end{figure}

\begin{figure}[H]
		\centering
		\includegraphics[scale=0.39]{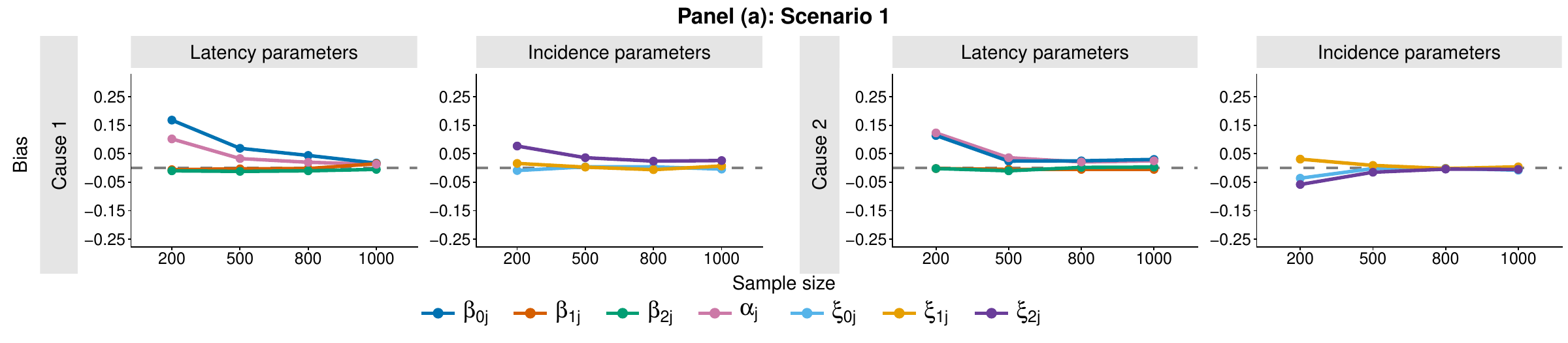}\\
		\includegraphics[scale=0.39]{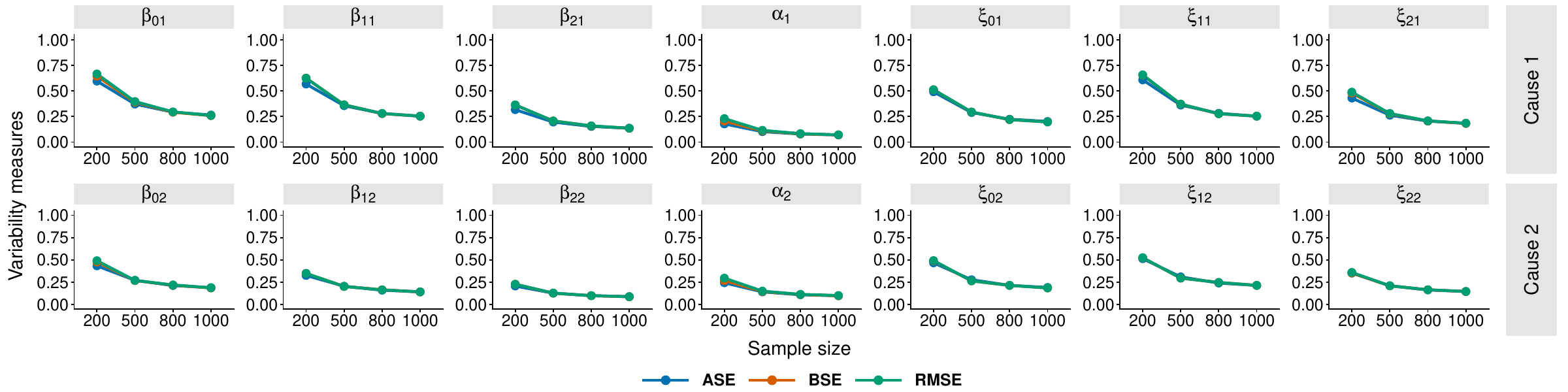}\\
		\includegraphics[scale=0.39]{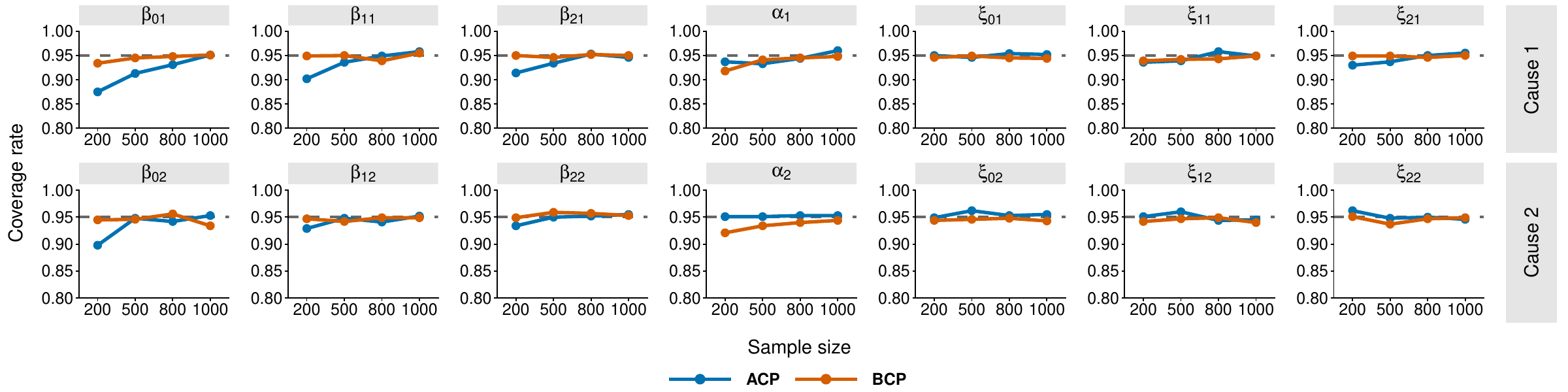}\\
		\hspace{0.01cm}
		\includegraphics[scale=0.39]{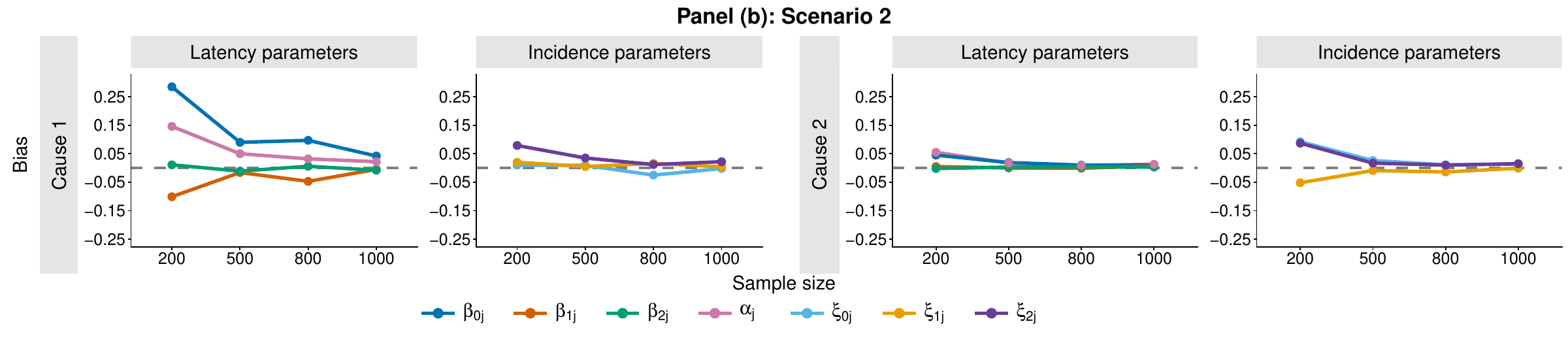}\\
		\includegraphics[scale=0.39]{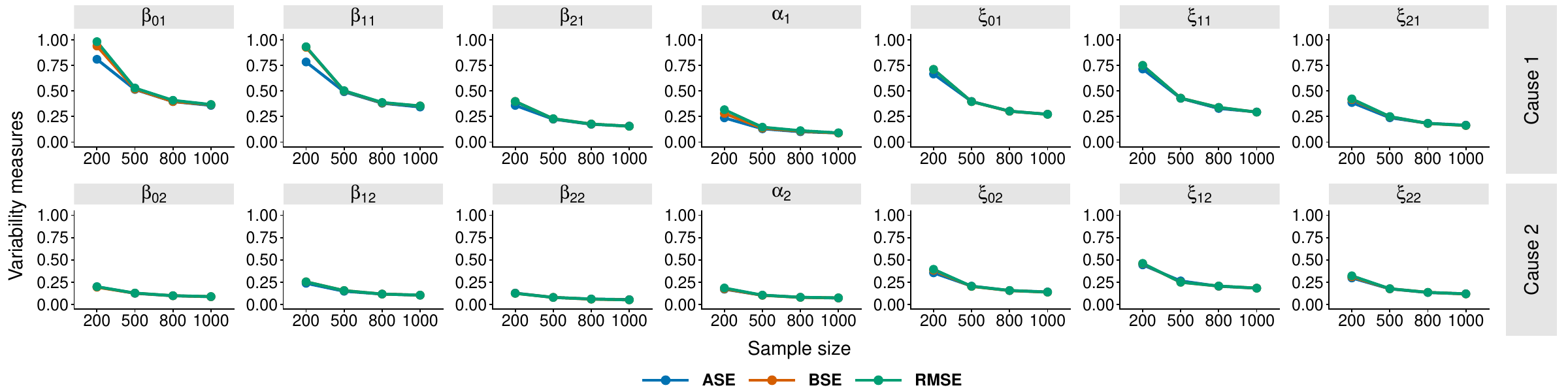}\\
		\includegraphics[scale=0.39]{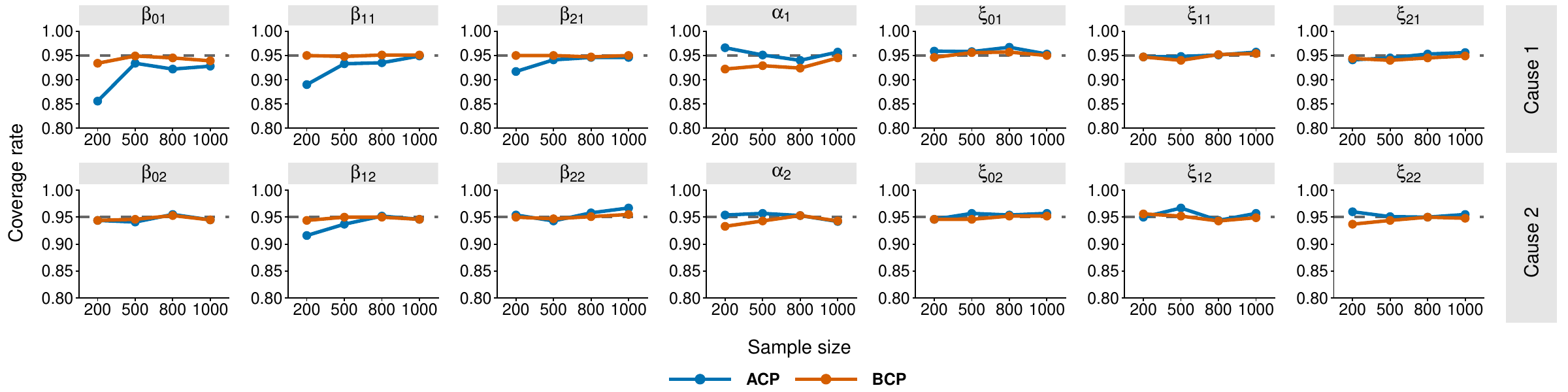}\\
		\vspace{-0.2cm}
		\caption{Monte Carlo simulation results for the Gamma-Multinomial mixture cure rate regression model under both scenarios, for $j=1$ (cause 1) and $j=2$ (cause 2).}
		\label{SimuGAMMA}
	\end{figure}

	The results of the Monte Carlo simulation study for the three finite-mixture cure rate models, evaluated under both simulation scenarios and across the two competing event types, are summarized in Figures~\ref{SimuWB}--\ref{SimuGAMMA}, with the corresponding numerical values provided in Tables~\ref{SimuTab.WB}--\ref{SimuTab.LogL} of the Supplementary Information. The models demonstrate satisfactory finite-sample performance in both scenarios and across competing causes, with estimation accuracy improving as the sample size increases. In particular, the maximum likelihood estimates obtained under the log-logistic and Gamma mixture cure rate models exhibit relatively large bias at $n = 200$; however, this bias decreases with increasing sample size, approaching zero for larger samples. Overall, the estimates show small bias across both the incidence and latency components, under both simulation scenarios and for all parametric latency distributions considered.
	
	The three measures of estimation uncertainty, ASE, BSE, and RMSE, display similar values and trends across all parameters. Although minor differences exist among these measures, particularly for BSE, these discrepancies diminish as the sample size increases, with all three measures approaching zero. A notable exception concerns the estimator of $\alpha_2$ in Scenario 1 under the Weibull model (Figure~\ref{SimuWB}), for which the empirical asymptotic and bootstrap coverage probabilities exhibit a distinct pattern. In particular, confidence intervals based on the BSE tend to be more conservative than those constructed using the ASE. For the remaining parameters, however, both ACP and BCP approach their nominal 95\% level as the sample size increases, across both competing causes and the incidence and latency components.
	
	Overall, the simulation results indicate that the proposed models perform reliably across all evaluated scenarios. The small empirical biases, decreasing estimation errors as sample size increases, greater agreement among error measures, and convergence of empirical coverage probabilities toward nominal levels collectively support the robustness of the estimation procedure. These findings are consistent across the Weibull, Gamma, and Log-Logistic latency distributions, suggesting that the proposed models perform satisfactorily under the parametric specifications considered. 
	
	\newpage
	
	\newpage
	
	\newpage
	
	\section{Data analysis}\label{sec4}
	
	The data analyzed in this study were obtained from the SEER program, as described in Section~\ref{sec1}, and include patients diagnosed with melanoma between 2003 and 2023. The available variables comprise sociodemographic characteristics (age at diagnosis, sex, and marital status) and clinical features (ulceration status, mitotic activity, histological type, lymph node involvement, and tumor stage). The outcome of interest was the time from diagnosis to death from melanoma. Deaths from other causes were treated as competing events, whereas patients who were alive at the end of follow-up were considered right-censored. The SEER dataset is ideal for demonstrating the proposed mixture cure model in the presence of competing risks. It allows for a comprehensive assessment of the probability of cure, the time until melanoma-related deaths, and the risk of dying from other causes. The analysis adopts a competing-risks framework, distinguishing between deaths due to melanoma and deaths from all other causes. This latter group includes fatalities from other cancers, a range of cardiovascular diseases (like heart attack, heart failure, or stroke), chronic respiratory illnesses, infections, metabolic conditions, pneumonia, and external factors such as accidents, suicide, or homicide, as well as any additional non-melanoma-related causes.
	
	After data cleaning and selection of relevant variables, a final cohort of 355 patients was identified. Among these, $n$=161 (45.35\%) died from melanoma, $n$=98 (27.61\%) died from other causes, and $n$=96 (27.04\%) were censored. The median failure time and interquartile range (IQR) were 1.17 years (IQR: 1.75 years) for death due to melanoma and 2.71 years (IQR: 5.06 years) for death from other causes. To improve patient survival and quality of life, the overall descriptive analysis of clinical and sociodemographic characteristics is as follows: sex (male: 74.53$\%$, female: 25.47$\%$); age group at diagnosis ($<$60 years: 26.71$\%$, $\geq$ 60 years: 73.29$\%$); marital status (married: 49.69$\%$, unmarried: 50.31$\%$); ulceration status (present: 83.23$\%$, absent: 16.77$\%$); mitotic activity (present: 93.79$\%$, absent: 6.21$\%$); histological type (superficial and lentigo maligna: 9.32\%, nodular and acral lentiginous: 50.31$\%$, other types: 40.37$\%$); lymph node level (levels I and II: 33.54$\%$, level III: 18.63$\%$, other levels: 47.82$\%$); and tumor stage (stage I: 3.73$\%$, stage II and III: 68.32$\%$, stage IV: 27.95$\%$). Lymph node involvement and tumor stage were classified based on the American Joint Committee on Cancer (AJCC) 6th edition criteria \citep{SEER2002}. Specifically, levels I and II indicate early and intermediate regional lymph node involvement, respectively, while level III denotes matted regional lymph nodes, reflecting more advanced regional disease.
	
\begin{figure}[H]
		\centering
		\includegraphics[width=1\linewidth]{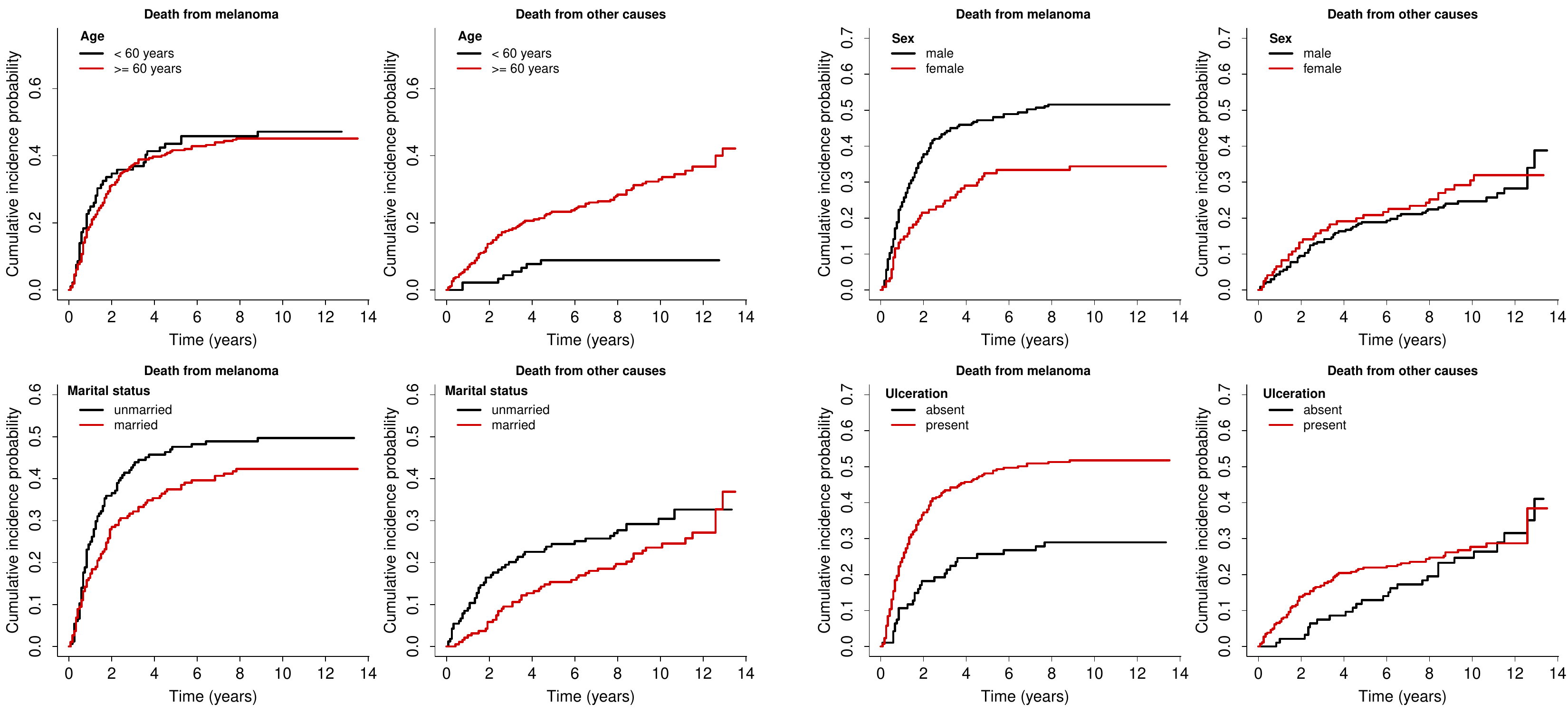}\\
		\includegraphics[width=1\linewidth]{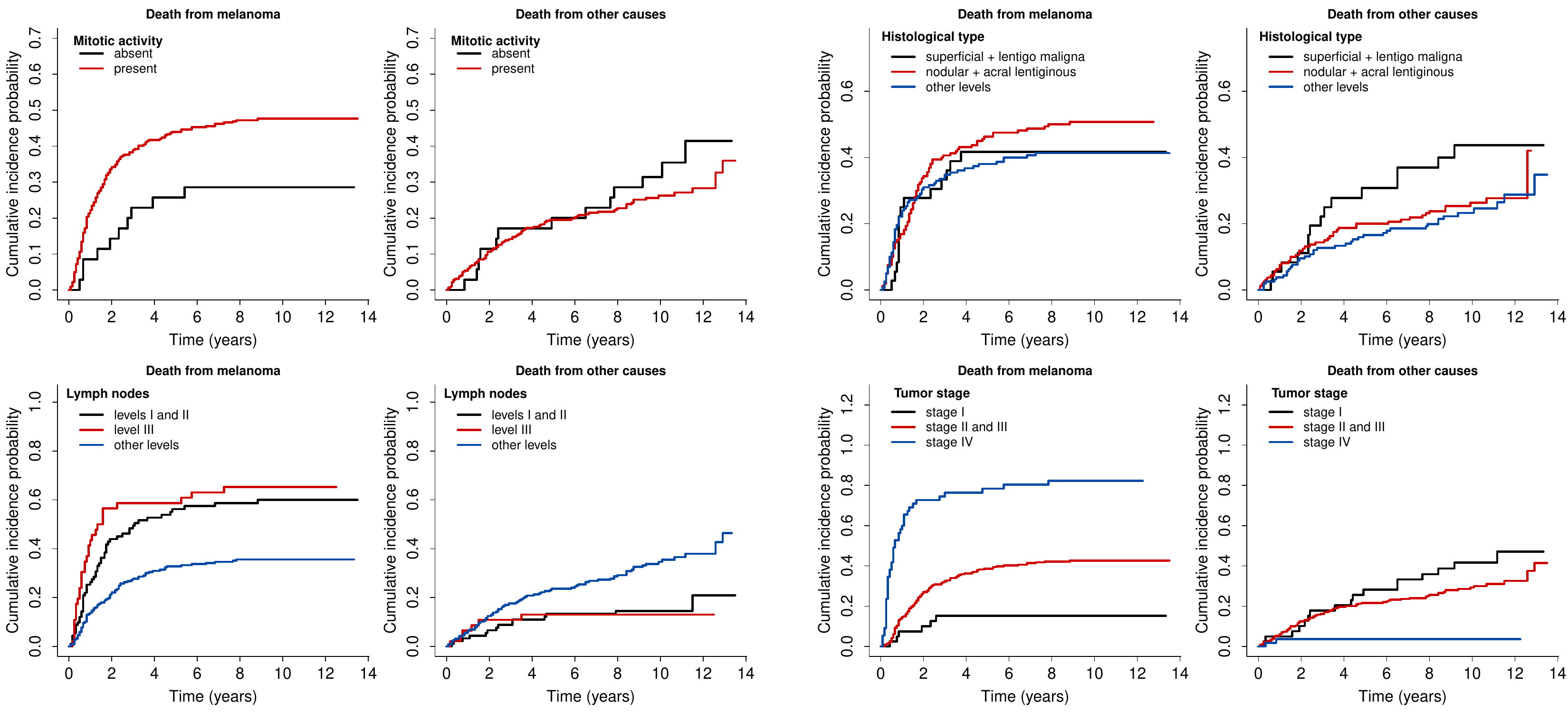}\\
		\vspace{-0.2cm}
		\caption{Estimated cumulative incidence function curves stratified by the demographic and clinical characteristics evaluated in the study.}
		\label{CIFs}
	\end{figure}

	The cumulative incidence probability plots presented in Figure~\ref{CIFs} provide compelling evidence that male patients face a significantly higher risk of melanoma-specific mortality compared to their female counterparts. Marital status also emerges as an important prognostic indicator for both causes of failure, with unmarried individuals (including those who are single, dating, widowed, or divorced) exhibiting substantially higher cumulative probabilities of death. In addition, the absence of ulceration or mitotic activity is strongly associated with a lower risk of melanoma-related death, highlighting their roles as protective factors.
	
	Additionally, clinical characteristics, particularly lymph node involvement and tumor stage, are strongly associated with the cumulative probability of melanoma-related death, emphasizing their significance in disease progression. In contrast, deaths from other causes, which comprise a heterogeneous group of events, exhibit substantial temporal variation in their cumulative incidence curves. Consequently, the probability of failure within a given covariate category may vary across follow-up intervals, reflecting the time-dependent nature of competing risks.

	To examine associations between risk factors and both the probability of cure and the occurrence of each type of failure in the melanoma dataset, we compared Weibull, Gamma, and Log-Logistic regression models. Model adequacy was evaluated using the maximized log-likelihood (LogL), Akaike information criterion (AIC), corrected AIC (AICc), Bayesian information criterion (BIC), and graphical comparisons of the fitted population survival and cumulative incidence functions against their empirical counterparts. As shown in Table~\ref{InfoCriteria} and in Figure~\ref{FigModelchoose} of Appendix~\ref{AdppA2}, the Log-Logistic mixture cure rate model provided the best overall fit and was selected for the subsequent analysis of covariate effects.
	
	\vspace{0.2cm}
	\begin{table}[H]
		\raggedleft
		\begin{threeparttable}
			\vspace{-0.4cm}
			\captionsetup{
				font={sf},
				justification=raggedright,
				singlelinecheck=false}
			\caption{\footnotesize Comparison of information criteria for the fitted parametric mixture cure rate models with competing risks.}
			\label{InfoCriteria}
			\setlength{\tabcolsep}{0.2cm}
			{\sf
				\footnotesize
				\begin{tabular}{lcccc}
					\toprule
					Regression model & LogL & AIC & AICc & BIC \\
					\midrule
					Weibull & -759.875& 1619.750&1636.363& 1813.777\\
					Gamma & -750.475 &1600.949& 1617.726&1794.555\\
					Log-logistic &-737.189&1574.378 & 1591.155 &1767.984 \\
					\bottomrule
				\end{tabular}
			}
		\end{threeparttable}
	\end{table}

	\begin{landscape}
		\begin{table}[H]
			\centering
			\sf
			\footnotesize
			\setlength{\tabcolsep}{0.15cm}
			\renewcommand{\arraystretch}{1.30}
			\caption{\sf \footnotesize Maximum likelihood estimates and their asymptotic and bootstrap standard errors for the incidence and latency components of the log-logistic mixture cure rate model with competing risks. The superscript $^{\dagger}$ indicates that the corresponding risk factor is statistically significant according to either the ASE or the BSE.}
			\label{TabResults}		
			\vspace{-0.2cm}
			\begin{tabular}{lrrrrrrrr}
				\hline
				& \multicolumn{4}{c}{\textbf{Latency distribution}}
				& \multicolumn{4}{c}{\textbf{Incidence distribution}}\\
				
				\cline{2-9}
				
				\textbf{Risk factor} & \multicolumn{2}{c}{Death from melanoma} & \multicolumn{2}{c}{Death from other causes} & \multicolumn{2}{c}{Death from melanoma} & \multicolumn{2}{c}{Death from other causes}\\
				\cline{2-9}
				
				& Estimate & ASE\,(BSE) & Estimate & ASE\, (BSE) & Estimate & ASE\, (BSE) & Estimate & ASE\, (BSE)\\
				
				\hline
				Intercept & 0.790 & 0.618 \,(0.712) & 1.647 & 0.424\, (0.416) & 0.039 & 0.984\, (0.757)	& 1.583 & 1.126\, (0.731)\\
				
				Age group ($<$ 60 years) & 0.176 & 0.181\, (0.208) & 0.181 & 0.489\, (0.379) & -1.194 & 0.416$^\dagger$\, (0.405$^\dagger$) & -2.092 & 0.646$^\dagger$\, (0.465$^\dagger$)\\
				
				Sex (female) & -0.024 & 0.192\, (0.228) & 0.390 & 0.256\, (0.341) & -1.172 & 0.423$^\dagger$\, (0.492$^\dagger$) & -0.769 & 0.503\, (0.531)\\
				
				Marital status (married) & 0.229 & 0.154\, (0.153) & 0.892 & 0.287$^\dagger$\, (0.349$^\dagger$) & -0.304 & 0.430\, (0.481) & -0.214 & 0.530\, (0.573)\\
				
				Ulceration (present) & -0.478 & 0.209$^\dagger$\, (0.204$^\dagger$) & -0.762 & 0.327$^\dagger$\, (0.307$^\dagger$) & 0.587 & 0.582\, (0.564) & -0.077 & 0.743\, (0.680)\\
				
				Mitotic activity (present)& -0.494 & 0.363\, (0.476) & -0.190 & 0.353\, (0.346) & -0.082 & 0.698\, (0.657)& 0.011 & 0.769\, (0.625)\\
				
				\multicolumn{9}{l}{\textit{Histological type (ref.: other levels)}}\\
				
				\hspace{0.2cm}superficial and lentigo maligna	& 0.245 & 0.286\, (0.316)	& -0.670 & 0.365\, (0.405) & 0.772 & 0.897\, (0.842) & 0.923 & 0.995\, (0.982)\\
				
				\hspace{0.2cm} nodular and acral lentiginous & 0.409 & 0.168$^\dagger$\, (0.173$^\dagger$) & -0.607 & 0.291$^\dagger$\, (0.364)& 0.501 & 0.477\, (0.550) & 0.318 & 0.592\, (0.677)\\
				
				\multicolumn{9}{l}{\textit{Lymph nodes (ref.: other levels)}}\\
				\hspace{0.2cm}
				levels I and II & -0.181 & 0.168\, (0.190) & -0.254 & 0.325\, (0.331) & 0.096 & 0.476\, (0.520) & -1.017 & 0.583\, (0.639)\\
				
				\hspace{0.2cm}
				level III      & -0.676 & 0.201$^\dagger$\, (0.201$^\dagger$) & -1.211 & 0.451$^\dagger$\, (0.457$^\dagger$) & 0.073 & 0.535\, (0.575) & -1.315 & 0.691\, (0.702)\\
				
				\multicolumn{9}{l}{\textit{Tumor stage (ref.: Level I)}}\\
				\hspace{0.2cm}
				stage II and III & 0.329 & 0.398\, (0.379) & 0.201 & 0.370\, (0.380) & 1.213 & 0.793\, (0.550$^\dagger$) & 0.425 & 0.837\, (0.521)\\
				
				\hspace{0.2cm}
				stage IV& -0.690 & 0.418\, (0.396) & -2.366 & 0.727$^\dagger$\, (0.596$^\dagger$) & 2.025 & 0.880$^\dagger$\, (0.614$^\dagger$) & -1.417 & 1.119\, (0.733)\\
				
				\hline
				$\hat{\alpha}_1$	& 1.924 & 0.148 (0.167)& & & & & & \\
				
				$\hat{\alpha}_2$ & & & 1.718 & 0.178 (0.186) & & & & \\
				\hline
			\end{tabular}
		\end{table}
	\end{landscape}
	
	The main results of the melanoma data analysis are summarized in Table~\ref{TabResults}. For melanoma-related mortality, significant risk factors at the 5\% level included ulceration ($p$-value $\approx 0.022$), nodular and acral lentiginous histological subtypes ($p$-value $\approx 0.015$), and level III lymph node involvement ($p$-value $<0.001$). For deaths from other causes, significant associations were observed with ulceration ($p$-value $\approx 0.013$), nodular and acral lentiginous histological subtypes ($p$-value $\approx 0.037$), marital status ($p$-value $\approx 0.002$), level III lymph node involvement ($p$-value $\approx 0.007$), and stage IV disease ($p$-value $<0.001$).
	
	It is important to note that certain clinical markers, particularly tumor stage and the presence of ulceration, would not typically be expected to reach statistical significance in the latency component for other-cause mortality, as opposed to melanoma-related mortality. This unexpected finding may reflect the complex clinical profiles of patients with advanced melanoma, such as increased disease burden, comorbidities, and exposure to more intensive treatments. Alternatively, the observed associations could stem, at least in part, from unmeasured patient characteristics or other sources of model uncertainty. These associations may also partially reflect differences in patient profiles not captured by the model. As such, these results should be interpreted within the context of the model specification and the available clinical information. The statistical significance of these covariates suggests an association with the timing of death from other causes among individuals susceptible to this event. Although the underlying mechanisms cannot be determined from the current analysis, these associations may signal the influence of complex clinical or patient-related factors not fully accounted for in the model. Further investigation, including more detailed clinical data, is needed to clarify these relationships.
	
	The estimated coefficients indicate that patients younger than 60 years, female, and married have lower susceptibility to experience both event types. In contrast, individuals with ulceration and mitotic activity, specific histological subtypes, lymph node involvement, or stage IV tumors demonstrate substantially higher susceptibility to death from melanoma and/or other causes. Based on the fitted model, the overall covariate-adjusted cure rate estimate is $\hat{\pi}_0\left(\bm{\xi}\right) = 0.186$, with a 95\% confidence interval of $(0.122; 0.249)$.

	To better understand how patient characteristics affect the long- and short-term survival, we created prognostic profiles that include both sociodemographic and clinical factors. For each profile, we calculated the likelihood of each of the two competing events occurring within 10 years, as well as the estimated probability of cure. We considered two sociodemographic scenarios representing less favorable and more favorable patient characteristics. In the less favorable scenario, patients were assumed to be aged 60 years or older, male, and unmarried. In the more favorable scenario, patients were assumed to be younger than 60 years and female, while marital status remained unchanged. Within each scenario, three clinical risk profiles were defined: low-risk, with all clinical characteristics set to their reference levels; intermediate-risk, characterized by the absence of ulceration but the presence of mitotic activity, a superficial spreading or lentigo maligna histological subtype, level I or II lymph node involvement, and stage II or III disease; and high-risk, characterized by the presence of both ulceration and mitotic activity, a nodular or acral lentiginous histological subtype, level III lymph node involvement, and stage IV disease. Keeping the clinical profiles unchanged across the two scenarios allowed us to examine differences in short- and long-term prognosis associated with the specified sociodemographic characteristics.
	
	\begin{table}[H]
		\centering
		\sf
		\footnotesize
		\caption{Estimated cure probabilities, cumulative incidence probabilities for melanoma-related and other-cause mortality, and population survival probabilities over time for low-, intermediate-, and high-risk clinical profiles under both less favorable and more favorable prognostic scenarios. The quantities $\hat{F}_j(t)$ and $\hat{\pi}_j(\bm{\xi})$ represent the estimated cumulative incidence function and the proportions of patients susceptible to death from melanoma ($j=1$) and other causes ($j=2$), respectively.}
		\vspace{-0.2cm}
		\label{TabProfile}
		\setlength{\tabcolsep}{0.15cm}
		\begin{tabular}{cccccccc}
			\hline
			\multicolumn{8}{c}{\textbf{More favorable prognostic scenario}}\\
			\hline
			Clinical risk profile & Time point (years) & $\hat{F}_1(t)$ &  $\hat{F}_2(t)$ & $\hat{S}_{pop}(t)$ &  $\hat{\pi}_1(\bm{\xi})$  &        $\hat{\pi}_2(\bm{\xi})$ & \text{Predicted cure fraction}\\ 
			\hline
			\multirow{4}{*}{Low} 
			& 1  & 0.005& 0.001& 0.994& \multirow{4}{*}{0.288} & \multirow{4}{*}{0.171} & \multirow{4}{*}{0.771} \\ 
			& 5  & 0.039& 0.012& 0.949&                  &                   &                   \\  
			& 8  & 0.047& 0.025& 0.927&                  &                   &                   \\  
			& 10 & 0.050& 0.035& 0.915&                  &                   &                   \\ 
			\hline
			\multirow{4}{*}{Intermediate} 
			& 1  & 0.032& 0.004& 0.964& \multirow{4}{*}{0.317} & \multirow{4}{*}{0.094} & \multirow{4}{*}{0.541} \\ 
			& 5  & 0.212& 0.046& 0.742&                   &                   &                   \\  
			& 8  & 0.251& 0.077& 0.671&                   &                   &                   \\  
			& 10 & 0.263& 0.093& 0.644&                   &                   &                   \\ 
			\hline
			\multirow{4}{*}{High} 
			& 1  & 0.496& 0.007& 0.497& \multirow{4}{*}{0.612} & \multirow{4}{*}{0.007} & \multirow{4}{*}{0.381} \\ 
			& 5  & 0.605& 0.007& 0.388&                   &                   &                   \\  
			& 8  & 0.609& 0.007& 0.384&                   &                   &                   \\  
			& 10 & 0.610& 0.007& 0.383&                   &                   &                   \\ 
			\hline
			\multicolumn{8}{c}{\textbf{Less favorable prognostic scenario}}\\
			\hline
			Clinical risk profile & Time point (years) & $\hat{F}_1(t)$ &  $\hat{F}_2(t)$ & $\hat{S}_{pop}(t)$ &  $\hat{\pi}_1(\bm{\xi})$  &        $\hat{\pi}_2(\bm{\xi})$ & \text{Predicted cure fraction}\\ 
			\hline
			
			\multirow{4}{*}{Low} 
			& 1  & 0.027& 0.039& 0.934& \multirow{4}{*}{0.150} & \multirow{4}{*}{0.705} & \multirow{4}{*}{0.145} \\ 
			& 5  & 0.125& 0.341& 0.534&                  &                   &                   \\  
			& 8  & 0.139& 0.477& 0.384&                  &                   &                   \\  
			& 10 & 0.143& 0.532& 0.325&                  &                   &                   \\ 
			\hline
			\multirow{4}{*}{Intermediate} 
			& 1  & 0.103& 0.098& 0.799& \multirow{4}{*}{0.494} & \multirow{4}{*}{0.442} & \multirow{4}{*}{0.064} \\ 
			& 5  & 0.422& 0.361& 0.217&                   &                   &                   \\  
			& 8  & 0.462& 0.402& 0.136&                   &                   &                   \\  
			& 10 & 0.473& 0.414& 0.114&                   &                   &                   \\ 
			\hline
			\multirow{4}{*}{High} 
			& 1  & 0.848& 0.017& 0.136& \multirow{4}{*}{0.943} & \multirow{4}{*}{0.017} & \multirow{4}{*}{0.041} \\ 
			& 5  & 0.938& 0.017& 0.045&                   &                   &                   \\  
			& 8  & 0.941& 0.017& 0.043&                   &                   &                   \\  
			& 10 & 0.941& 0.017& 0.042&                   &                   &                   \\ 
			\hline
		\end{tabular}
	\end{table}

\begin{figure}[H]
		\centering
		\includegraphics[scale=0.40]{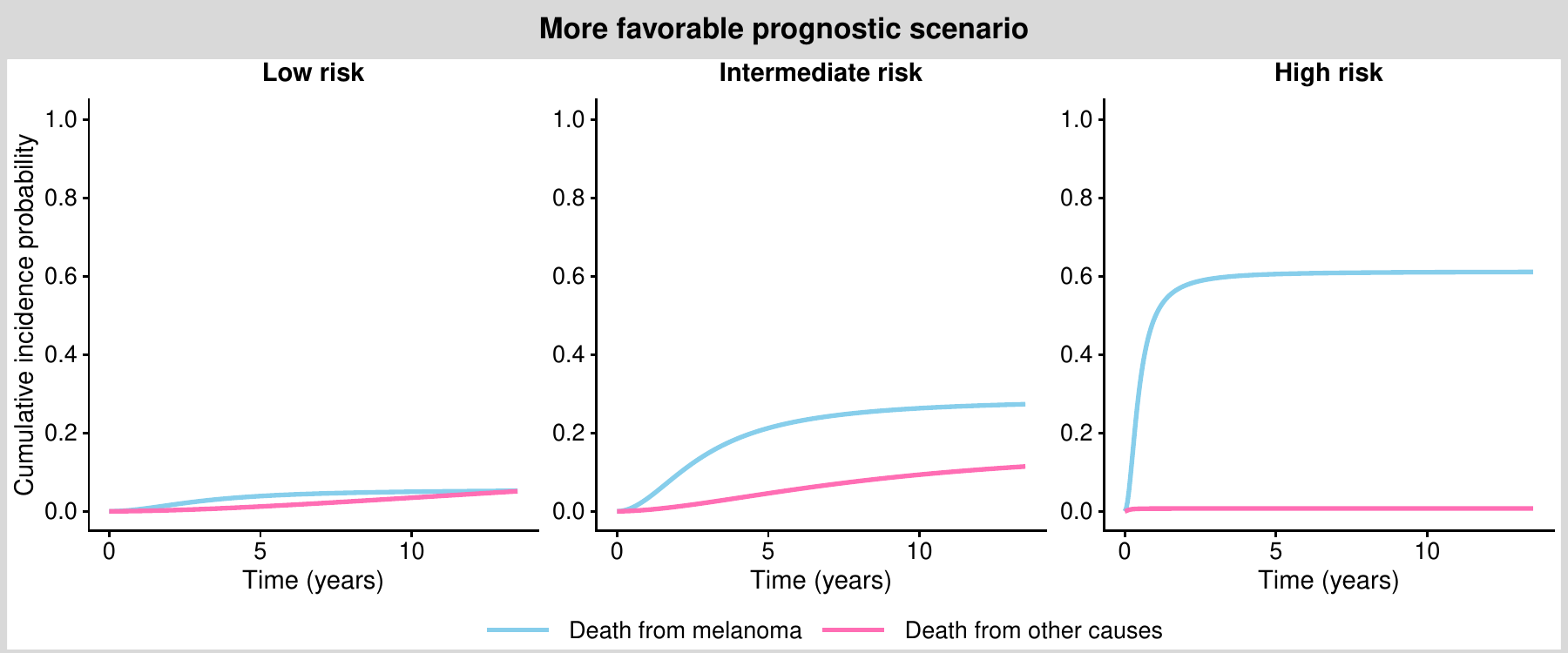}\\
		\includegraphics[scale=0.40]{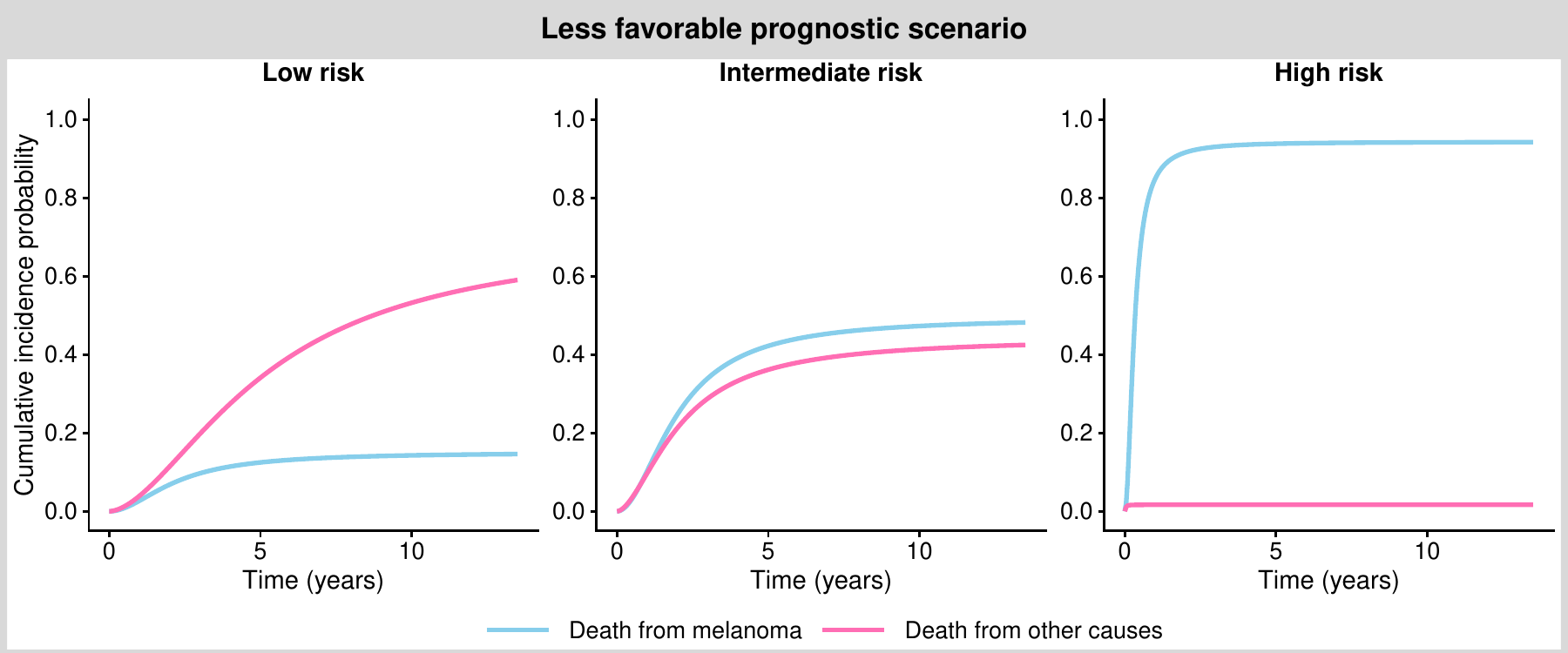}\\
		\caption{Estimated population-based cumulative incidence functions for melanoma-related and other-cause mortality, stratified by low-, intermediate-, and high-risk clinical profiles.}
		\label{FigPerfis}
	\end{figure}

	Figure~\ref{FigPerfis} illustrates distinct patterns in the cumulative incidence of death from the two competing causes across clinical risk profiles. Under the more favorable prognostic scenario, both causes exhibit low cumulative incidence in the low-risk group, with similar probabilities of death from melanoma and other causes. In the intermediate-risk group, melanoma-related mortality becomes more prominent. This trend is even more pronounced in the high-risk group, where the cumulative incidence of melanoma-related death increases rapidly while deaths from other causes remain close to zero. A similar pattern emerges under the less favorable sociodemographic scenario, although the overall probabilities of death are substantially higher. In the low-risk group, death from other causes is more likely than melanoma-related death. In the intermediate-risk group, the cumulative incidence curves for the two causes become closer, although melanoma-related mortality remains higher. In the high-risk group, the cumulative incidence of melanoma-related death approaches one, indicating that it is the predominant cause of death predicted by the model for this profile. Overall, these findings highlight the differences in cause-specific mortality across clinical risk profiles and suggest that more favorable prognostic characteristics are associated with a greater probability of long-term survival.
	
	These findings align with the results in Table~\ref{TabProfile}, which analyzes the temporal dynamics of the three clinical risk profiles across both prognostic scenarios. The table elucidates changes in survival, cure, and cause-specific mortality probabilities at specific time points. In the more favorable prognostic scenario, patients exhibit substantially higher predicted cure fractions: 77.1\% for the low-risk profile, 54.1\% for the intermediate-risk profile, and 38.1\% for the high-risk profile. Across the temporal horizon, melanoma-related mortality consistently shows the highest cumulative incidence among patients with intermediate- and high-risk profiles, irrespective of the prognostic scenario. In contrast, for patients with a low-risk profile under the less favorable scenario, the cumulative probability of death from other causes exceeds that of melanoma-related death. Nevertheless, melanoma remains the predominant cause of mortality throughout the period for intermediate- and high-risk groups. Notably, in the less favorable prognostic scenario, individuals with a high-risk profile experience a cumulative probability of melanoma-related death that increases from 84.8\% at one year to 94.1\% at ten years. The estimated cure fraction also decreases markedly across the clinical risk profiles, from 14.4\% in the low-risk group to just 4.1\% in the high-risk group. These findings further highlight the substantial differences in long-term survival across risk profiles.
	
	This profile-based analysis is limited to the covariates included in the model. Other factors, such as treatment type, may also influence prognosis but were not considered because of the high proportion of missing data.
	
\begin{figure}[H]
		\raggedleft
		\begin{threeparttable}
			\captionsetup{
				font={sf},
				justification=raggedright,
				singlelinecheck=false}
			\includegraphics[width=0.6\linewidth]{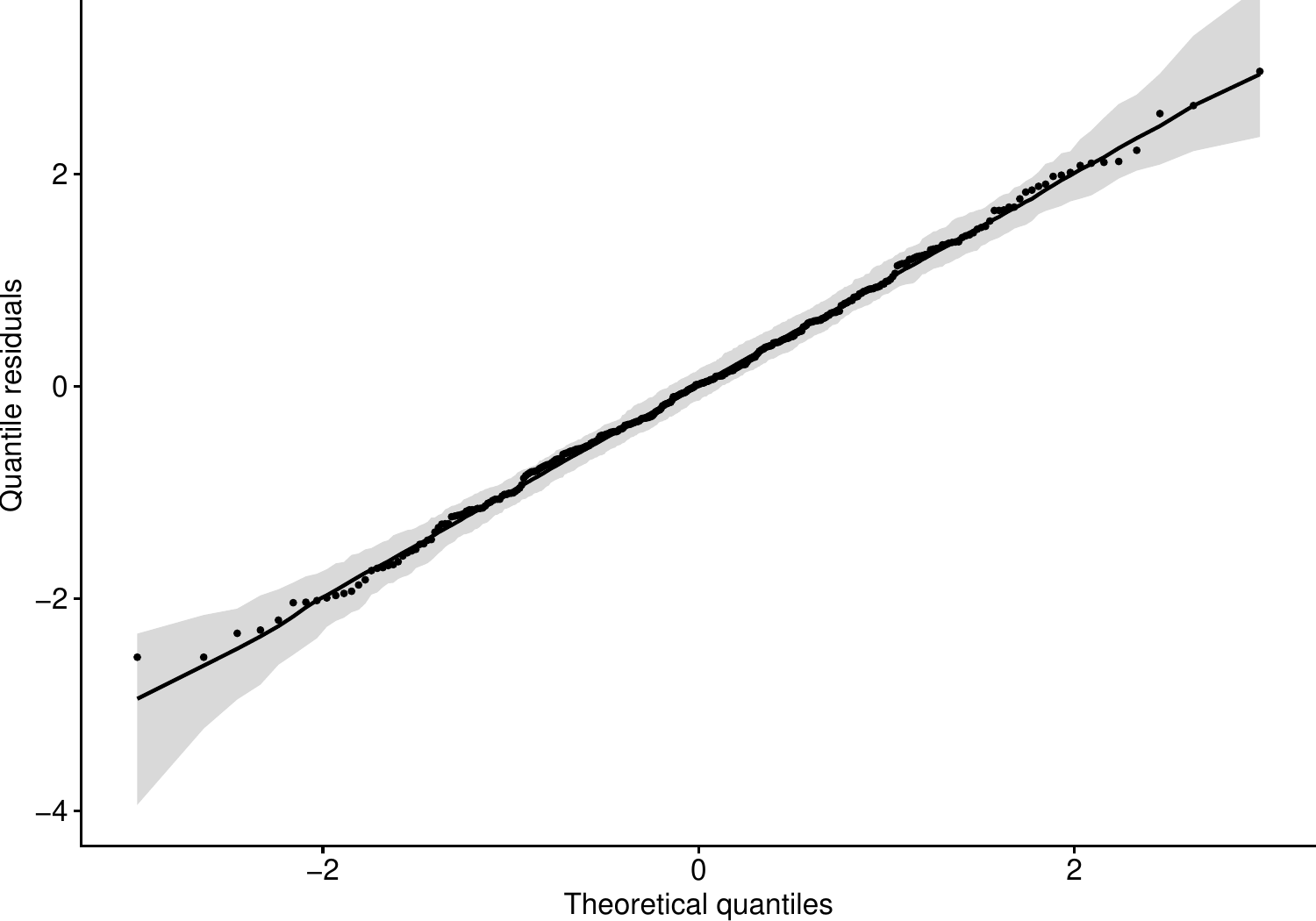}
		\end{threeparttable}
		\caption{Quantile residual plots for the Log-logistic-multinomial mixture model with competing risks.}
		\label{Residuals}
	\end{figure}

	The quantile–quantile plot in Figure~\ref{Residuals} shows good agreement between the empirical and theoretical residual quantiles, with most points lying close to the reference line. Although some deviations are apparent in the tails, most remain within the confidence band. Overall, the plot supports the adequacy of the assumed residual distribution and provides graphical evidence of a satisfactory model fit.
	
	Overall, our findings are consistent with previous studies in the oncology literature \citep{retsas2002prognostic,chiu2026clinical}, which highlight the prognostic relevance of clinical and pathological features in patients with primary cutaneous melanoma. In particular, these factors are associated with time to melanoma-related death, with certain characteristics linked to more rapid disease progression and poorer prognosis.

	\section{Final remarks} \label{sec5}
	This paper proposes a parametric mixture cure model for competing risks that accommodates long-term survivors and multiple causes of failure. Model parameters were estimated by maximum likelihood, facilitated by the expectation–maximization algorithm. Extensive simulation studies assess finite-sample performance in terms of empirical bias, root mean squared error, concordance between bootstrap and asymptotic standard errors, and coverage probabilities. The results indicate that the proposed estimation procedure performs consistently across the scenarios considered. The proposed methodology was applied to melanoma patient data from the SEER program. The findings highlight the importance of jointly modeling the cure fraction and competing risks in cancer survival analysis and identify risk factors associated with cause-specific susceptibility and event-time distributions among susceptible individuals. Residual diagnostics confirm the adequacy of the model fit, and the results are consistent with previous research in dermatologic oncology. Overall, the proposed model offers a flexible framework for competing risks survival analysis by jointly characterizing the cure fraction, cause-specific susceptibility, and event-time distributions. Its application to other clinical and epidemiological settings is promising, provided that model assumptions are rigorously assessed and that available data sufficiently support the identification of both the cure fraction and cause-specific components.

	\subsection*{Author Contributions}
	All authors contributed equally to the conception and design of the study, data analysis, interpretation of the results, manuscript preparation, and revision. All authors read and approved the final version of the manuscript.

	\section*{Acknowledgment}
	The first and second authors acknowledge financial support from Coordenação de Aperfeiçoamento de Pessoal de Nível Superior (CAPES), Brazil, and the Decanato de Pós-Graduação (DPG) at the University of Brasília, respectively. The authors thank the Surveillance, Epidemiology, and End Results Program for providing access to its database. The findings and conclusions are those of the authors and do not necessarily reflect the views of the National Cancer Institute.

	\subsection*{Financial Disclosure}
	None reported.
	
	\subsection*{Conflicts of Interest}
	The authors declare no conflicts of interest.
	
	\bibliographystyle{apalike}
	\bibliography{refbib}
	
	\section*{Appendices}

\setcounter{section}{0}
\renewcommand{\thesection}{A\arabic{section}}
\setcounter{equation}{0}
\renewcommand{\theequation}{A\arabic{equation}}
\counterwithout{figure}{section}
\setcounter{figure}{0}
\renewcommand{\thefigure}{A1.\arabic{figure}}
\counterwithout{table}{section}
\setcounter{table}{0}
\renewcommand{\thetable}{A1.\arabic{table}}
\section*{Additional Results}

This appendix presents supplementary results that support and extend the findings reported in the main paper and is organized into two sections: Section~\ref{AdppA1} provides additional simulation results, while Section~\ref{AdppA2} presents further results from the application of the proposed methodology to the melanoma dataset.

\newpage
\section{Additional simulation results}\label{AdppA1}
This section of the appendix presents the numerical results of the simulation study reported in the main text, providing additional details to support the findings.

\begin{table}[H]
\setlength{\tabcolsep}{0.10cm}
\setstretch{1.0}
\centering
\sf
\scriptsize
\caption{Monte Carlo simulation results for the Weibull-Multinomial regression model.}
\label{SimuTab.WB}
\vspace{-0.2cm}
\begin{tabular}{rrrrrrrrrrrrrrrr}
\hline
\multicolumn{16}{c}{\text{\bf Scenario 1}}\\
\hline
\multirow{2}{*}{} & \multicolumn{7}{c}{Cause 1}& \multicolumn{7}{c}{Cause 2}\\
\cline{2-8} \cline{10-16}
& $\beta_{01}$ & $\beta_{11}$ & $\beta_{21}$ & $\alpha_{1}$ & $\xi_{01}$ & $\xi_{11}$ & $\xi_{21}$ && $\beta_{02}$ & $\beta_{12}$ & $\beta_{22}$ & $\alpha_{2}$ & $\xi_{02}$ & $\xi_{12}$ & $\xi_{22}$ \\
\hline
\multicolumn{16}{c}{$n=200$}\\
\hline
Bias&-0.024&-0.013&0.001&0.041&-0.013&0.012&0.060&&0.000&-0.008&0.021&0.004&-0.010&0.025&-0.034\\
RMSE& 0.347& 0.393&0.252&0.101& 0.331&0.433&0.292&&0.218& 0.278&0.182&0.124& 0.305&0.449&0.327\\
ASE & 0.333& 0.369&0.239&0.093& 0.314&0.422&0.281&&0.229& 0.289&0.188&0.144& 0.313&0.452&0.314\\
BSE & 0.346& 0.392&0.251&0.092& 0.331&0.433&0.286&&0.218& 0.278&0.180&0.123& 0.305&0.448&0.325\\
ACP & 0.934& 0.931&0.930&0.957& 0.938&0.945&0.955&&0.960& 0.958&0.957&0.973& 0.968&0.956&0.945\\
BCP & 0.948& 0.944&0.942&0.929& 0.949&0.942&0.946&&0.942& 0.952&0.945&0.946& 0.953&0.949&0.947\\
\hline
\multicolumn{16}{c}{$n=500$}\\
\hline
Bias&-0.014&0.009&0.001&0.016&-0.008&-0.003&0.039&& 0.009&-0.013&0.014&-0.008&-0.002&0.010&-0.009\\
RMSE& 0.211&0.237&0.153&0.060& 0.193& 0.266&0.178&& 0.136& 0.173&0.110& 0.075&0.210&0.293&0.186\\
ASE & 0.211&0.233&0.148&0.057& 0.195& 0.262&0.172&& 0.142& 0.179&0.114& 0.090&0.194&0.279&0.192\\
BSE & 0.220&0.236&0.153&0.058& 0.193& 0.265&0.174&& 0.135& 0.172&0.109& 0.074&0.190&0.2293&0.186\\
ACP & 0.947&0.948&0.953&0.950& 0.956& 0.950&0.956&& 0.961& 0.958&0.956& 0.982&0.933&0.941&0.947\\
BCP & 0.955&0.960&0.953&0.939& 0.953& 0.950&0.945&& 0.952& 0.942&0.957& 0.950&0.945&0.948&0.939\\
\hline
\multicolumn{16}{c}{$n=800$}\\
\hline
Bias&-0.018&0.012&0.005&0.008&-0.010&0.000&0.035&&0.011&-0.004&0.014&-0.008&-0.005&0.007&-0.001\\
RMSE& 0.177&0.184&0.120&0.046& 0.155&0.207&0.143&&0.113& 0.143&0.085& 0.061& 0.158&0.219& 0.151\\
ASE & 0.167&0.185&0.117&0.044& 0.154&0.206&0.135&&0.112& 0.141&0.089& 0.071& 0.153&0.220& 0.151\\
BSE & 0.176&0.184&0.119&0.045& 0.154&0.207&0.138&&0.112& 0.142&0.084& 0.061& 0.158&0.218& 0.151\\
ACP & 0.941&0.960&0.952&0.952& 0.951&0.952&0.936&&0.943& 0.942&0.954& 0.977& 0.948&0.948& 0.948\\
BCP & 0.951&0.951&0.958&0.943& 0.943&0.954&0.938&&0.945& 0.944&0.942& 0.949& 0.956&0.947& 0.943\\
\hline
\multicolumn{16}{c}{$n=1,000$}\\
\hline
Bias&-0.001&0.005&-0.004&0.010&0.001&0.009&0.023&&0.008&-0.005&0.012&-0.014&0.06&-0.002&-0.002\\
RMSE& 0.151&0.165& 0.104&0.042&0.134&0.178&0.119&&0.098& 0.122&0.079& 0.056&0.130&0.194& 0.133\\
ASE & 0.149&0.165& 0.104&0.040&0.137&0.184&0.120&&0.098& 0.126&0.079& 0.063&0.136&0.196& 0.135\\
BSE & 0.141&0.165& 0.104&0.039&0.134&0.178&0.116&&0.095& 0.122&0.078& 0.054&0.130&0.194& 0.133\\
ACP & 0.940&0.942& 0.955&0.938&0.956&0.962&0.958&&0.957& 0.955&0.953& 0.965&0.963&0.956& 0.955\\
BCP & 0.951&0.943& 0.947&0.941&0.947&0.954&0.944&&0.950& 0.941&0.947& 0.949&0.953&0.947& 0.949\\
\hline
\multicolumn{16}{c}{\text{\bf Scenario 2}}\\
\hline
\multirow{2}{*}{} & \multicolumn{7}{c}{Cause 1}& \multicolumn{7}{c}{Cause 2}\\
\cline{2-8} \cline{10-16}
& $\beta_{01}$ & $\beta_{11}$ & $\beta_{21}$ & $\alpha_{1}$ & $\xi_{01}$ & $\xi_{11}$ & $\xi_{21}$ && $\beta_{02}$ & $\beta_{12}$ & $\beta_{22}$ & $\alpha_{2}$ & $\xi_{02}$ & $\xi_{12}$ & $\xi_{22}$ \\
\hline
\multicolumn{16}{c}{$n=200$}\\
\hline
Bias&-0.060&0.065&-0.060&0.078&0.000&0.055&0.011&&-0.005&-0.012&-0.003&0.046&0.079&-0.030&0.039\\
RMSE& 0.563&0.646& 0.396&0.166&0.576&0.701&0.373&& 0.199& 0.243& 0.153&0.135&0.482& 0.575&0.348\\
ASE & 0.535&0.614& 0.346&0.138&0.568&0.666&0.363&& 0.188& 0.230& 0.143&0.119&0.568& 0.666&0.363\\
BSE & 0.558&0.642& 0.390&0.146&0.576&0.698&0.373&& 0.199& 0.242& 0.153&0.127&0.476& 0.574&0.345\\
ACP & 0.935&0.926& 0.915&0.944&0.973&0.959&0.957&& 0.929& 0.930& 0.927&0.943&0.977& 0.965&0.956\\
BCP & 0.940&0.947& 0.945&0.926&0.950&0.947&0.953&& 0.943& 0.952& 0.943&0.938&0.935& 0.946&0.950\\
\hline
\multicolumn{16}{c}{$n=500$}\\
\hline
Bias&-0.021&-0.006&-0.007&0.025&-0.010&0.011&0.006&&-0.003&-0.001&-0.002&0.015&0.023&-0.024&0.029\\
RMSE& 0.346& 0.390& 0.225&0.087& 0.321&0.392&0.218&& 0.118& 0.146& 0.092&0.076&0.279& 0.339&0.203\\
ASE & 0.331& 0.380& 0.206&0.081& 0.328&0.388&0.213&& 0.119& 0.145& 0.090&0.073&0.282& 0.339&0.200\\
BSE & 0.345& 0.390& 0.225&0.083& 0.321&0.392&0.218&& 0.118& 0.146& 0.091&0.075&0.278& 0.338&0.201\\
ACP & 0.936& 0.953& 0.925&0.946& 0.962&0.949&0.946&& 0.939& 0.940& 0.951&0.955&0.955& 0.958&0.953\\
BCP & 0.951& 0.955& 0.950&0.939& 0.948&0.946&0.949&& 0.942& 0.947& 0.950&0.946&0.950& 0.951&0.944 \\
\hline
\multicolumn{16}{c}{$n=800$}\\
\hline
Bias&-0.020& 0.001&-0.005&0.019&0.014&-0.010&0.016&&-0.001&-0.002&-0.004&0.011&0.018&-0.015&0.016\\
RMSE& 0.273& 0.312& 0.165&0.069&0.251& 0.298&0.170&& 0.092& 0.112& 0.070&0.058&0.213& 0.260&0.155\\
ASE & 0.255& 0.293& 0.158&0.063&0.254& 0.301&0.167&& 0.094& 0.115& 0.071&0.058&0.219& 0.264&0.157\\
BSE & 0.272& 0.313& 0.165&0.066&0.251& 0.298&0.169&& 0.092& 0.112& 0.070&0.057&0.213& 0.259&0.154\\
ACP & 0.920& 0.932& 0.938&0.937&0.948& 0.953&0.950&& 0.955& 0.951& 0.951&0.962&0.966& 0.943&0.956\\
BCP & 0.941& 0.950& 0.950&0.935&0.944& 0.949&0.948&& 0.951& 0.943& 0.941&0.949&0.944& 0.939&0.949\\
\hline
\multicolumn{16}{c}{$n=1,000$}\\
\hline
Bias&-0.002&0.006&-0.014&0.014&-0.001&0.001&0.002&&-0.005&0.002&0.000&0.011&0.008&-0.004&0.004\\
RMSE& 0.238&0.269& 0.148&0.059& 0.221&0.262&0.145&& 0.083&0.103&0.064&0.051&0.191& 0.234&0.141\\
ASE & 0.231&0.264& 0.144&0.056& 0.226&0.269&0.149&& 0.083&0.102&0.063&0.052&0.195& 0.235&0.139\\
BSE & 0.238&0.269& 0.147&0.058& 0.220&0.262&0.145&& 0.083&0.103&0.064&0.050&0.191& 0.234&0.141\\
ACP & 0.947&0.949& 0.956&0.950& 0.961&0.963&0.966&& 0.950&0.955&0.946&0.953&0.953& 0.950&0.945\\
BCP & 0.951&0.954& 0.953&0.941& 0.958&0.958&0.964&& 0.954&0.952&0.956&0.940&0.951& 0.950&0.944\\
\hline
\end{tabular}
\end{table}

\newpage
\begin{table}[H]
\setlength{\tabcolsep}{0.10cm}
\setstretch{1.1}
\centering
\sf
\scriptsize
\caption{Monte Carlo simulation results for the Log-Logistic-Multinomial regression model.}
\label{SimuTab.LogL}
\vspace{-0.2cm}
\begin{tabular}{rrrrrrrrrrrrrrrr}
\hline
\multicolumn{16}{c}{\text{\bf Scenario 1}}\\
\hline
\multirow{2}{*}{} & \multicolumn{7}{c}{Cause 1}& \multicolumn{7}{c}{Cause 2}\\
\cline{2-8} \cline{10-16}
& $\beta_{01}$ & $\beta_{11}$ & $\beta_{21}$ & $\alpha_{1}$ & $\xi_{01}$ & $\xi_{11}$ & $\xi_{21}$ && $\beta_{02}$ & $\beta_{12}$ & $\beta_{22}$ & $\alpha_{2}$ & $\xi_{02}$ & $\xi_{12}$ & $\xi_{22}$ \\
\hline
\multicolumn{16}{c}{$n=200$}\\
\hline
Bias&0.168&-0.006&-0.010&0.102&-0.009&0.016&0.077&&0.114&-0.002&-0.002&0.123&-0.036&0.031&-0.058\\
RMSE&0.667& 0.626& 0.364&0.230& 0.513&0.658&0.489&&0.492& 0.350& 0.229&0.297& 0.493&0.526& 0.361\\
ASE &0.597& 0.568& 0.317&0.178& 0.492&0.609&0.431&&0.434& 0.327& 0.209&0.245& 0.468&0.514& 0.353\\
BSE &0.645& 0.626& 0.363&0.206& 0.512&0.656&0.480&&0.477& 0.350& 0.229&0.270& 0.490&0.524& 0.355\\
ACP &0.875& 0.902& 0.914&0.937& 0.950&0.936&0.930&&0.898& 0.929& 0.934&0.951& 0.949&0.951& 0.962\\
BCP &0.934& 0.949& 0.950&0.918& 0.946&0.939&0.949&&0.945& 0.947& 0.949&0.921& 0.944&0.942& 0.951\\
\hline
\multicolumn{16}{c}{$n=500$}\\
\hline
Bias&0.069&-0.003&-0.012&0.033&0.004&0.003&0.036&&0.024&-0.005&-0.010&0.036&-0.002&0.009&-0.015\\
RMSE&0.397& 0.365& 0.207&0.115&0.295&0.372&0.280&&0.271& 0.204& 0.130&0.152& 0.266&0.297& 0.211\\
ASE &0.372& 0.356& 0.196&0.102&0.289&0.362&0.263&&0.273& 0.207& 0.129&0.143& 0.279&0.311& 0.212\\
BSE &0.391& 0.365& 0.207&0.110&0.294&0.371&0.277&&0.270& 0.204& 0.129&0.148& 0.266&0.297& 0.211\\
ACP &0.913& 0.936& 0.934&0.933&0.946&0.939&0.937&&0.948& 0.948& 0.950&0.951& 0.962&0.960& 0.948\\
BCP &0.945& 0.950& 0.946&0.941&0.949&0.942&0.949&&0.946& 0.942&0.959& 0.934& 0.946&0.947& 0.937\\
\hline
\multicolumn{16}{c}{$n=800$}\\
\hline
Bias&0.044&-0.002&-0.010&0.020&0.004&-0.006&0.024&&0.025&-0.005&0.002&0.021&-0.001&-0.002&-0.004\\
RMSE&0.296& 0.280& 0.159&0.083&0.220& 0.277&0.208&&0.221& 0.166&0.101&0.116& 0.216& 0.249& 0.167\\
ASE &0.293& 0.280& 0.153&0.079&0.224& 0.282&0.205&&0.214& 0.162&0.101&0.111& 0.216& 0.242& 0.165\\
BSE &0.292& 0.279& 0.158&0.081&0.220& 0.277&0.207&&0.219& 0.167&0.101&0.114& 0.216& 0.249& 0.167\\
ACP &0.931& 0.949& 0.953&0.944&0.954& 0.958&0.950&&0.942& 0.941&0.952&0.953& 0.953& 0.944& 0.950\\
BCP &0.948& 0.939& 0.952&0.945&0.945& 0.943&0.946&&0.956& 0.949&0.957&0.940& 0.948& 0.949& 0.947\\
\hline
\multicolumn{16}{c}{$n=1,000$}\\
\hline
Bias&0.017&0.015&-0.005&0.013&-0.004&0.007&0.026&&0.030&-0.005&0.003&0.025&-0.008&0.004&-0.005\\
RMSE&0.260&0.254& 0.136&0.071& 0.197&0.254&0.185&&0.189& 0.143&0.090&0.103& 0.187&0.216& 0.148\\
ASE &0.265&0.253& 0.136&0.070& 0.202&0.253&0.183&&0.191& 0.144&0.090&0.100& 0.193&0.216& 0.147\\
BSE &0.259&0.254& 0.135&0.070& 0.197&0.254&0.183&&0.187& 0.143&0.089&0.100& 0.187&0.216& 0.148\\
ACP &0.951&0.958& 0.946&0.960& 0.952&0.949&0.955&&0.953& 0.952&0.955&0.953& 0.955&0.945& 0.946\\
BCP &0.951&0.955& 0.950&0.948& 0.944&0.949&0.950&&0.934& 0.949&0.953&0.944& 0.943&0.940& 0.949\\
\hline
\multicolumn{16}{c}{\text{\bf Scenario 2}}\\
\hline
\multirow{2}{*}{} & \multicolumn{7}{c}{Cause 1}& \multicolumn{7}{c}{Cause 2}\\
\cline{2-8} \cline{10-16}
& $\beta_{01}$ & $\beta_{11}$ & $\beta_{21}$ & $\alpha_{1}$ & $\xi_{01}$ & $\xi_{11}$ & $\xi_{21}$ && $\beta_{02}$ & $\beta_{12}$ & $\beta_{22}$ & $\alpha_{2}$ & $\xi_{02}$ & $\xi_{12}$ & $\xi_{22}$ \\
\hline
\multicolumn{16}{c}{$n=200$}\\
\hline
Bias&0.285&-0.101&0.011&0.146&0.010&0.020&0.079&&0.045&0.005&-0.002&0.055&0.092&-0.052&0.087\\
RMSE&0.983& 0.933&0.398&0.317&0.712&0.751&0.423&&0.201&0.256&0.127& 0.187&0.396& 0.462&0.323\\
ASE &0.810& 0.784&0.359&0.237&0.666&0.715&0.387&&0.201&0.239& 0.130&0.173&0.356& 0.445&0.299\\
BSE &0.940& 0.927&0.398&0.280&0.711&0.750&0.416&&0.195&0.256& 0.127&0.178&0.385& 0.459&0.312\\
ACP &0.856& 0.890&0.917&0.966&0.959&0.947&0.941&&0.944&0.916& 0.954&0.954&0.946& 0.950&0.960\\
BCP &0.934& 0.950&0.950&0.922&0.946&0.947&0.944&&0.944&0.944& 0.950&0.933&0.946& 0.956&0.937\\
\hline
\multicolumn{16}{c}{$n=500$}\\
\hline
Bias&0.090&-0.016&-0.011&0.050&0.009&0.005&0.035&&0.019&0.000&0.003&0.017&0.026&-0.009&0.017\\
RMSE&0.529& 0.503& 0.229&0.146&0.396&0.430&0.250&&0.130&0.158&0.082&0.107&0.206& 0.251&0.179\\
ASE &0.515& 0.493& 0.224&0.130&0.398&0.428&0.238&&0.127&0.150&0.081&0.106&0.208& 0.266&0.177\\
BSE &0.519& 0.501& 0.228&0.137&0.396&0.430&0.248&&0.129&0.158&0.082&0.105&0.204& 0.251&0.178\\
ACP &0.934& 0.933& 0.941&0.951&0.958&0.948&0.945&&0.941&0.937&0.943&0.957&0.957& 0.967&0.951\\
BCP &0.949& 0.948& 0.950&0.929&0.956&0.940&0.940&&0.946&0.950&0.947&0.943&0.946& 0.952&0.944\\
\hline
\multicolumn{16}{c}{$n=800$}\\
\hline
Bias&0.097&-0.047&0.006&0.032&-0.025&0.016&0.012&&0.010&-0.001&0.004&0.009&0.011&-0.014&0.010\\
RMSE&0.409& 0.388&0.177&0.112& 0.303&0.340&0.183&&0.101& 0.119&0.063&0.083&0.159& 0.209&0.138\\
ASE &0.396& 0.380&0.174&0.101& 0.302&0.329&0.184&&0.100& 0.119&0.064&0.083&0.158& 0.206&0.136\\
BSE &0.397& 0.384&0.177&0.107& 0.302&0.339&0.182&&0.100& 0.119&0.062&0.082&0.158& 0.208&0.138\\
ACP &0.922& 0.935&0.946&0.940& 0.967&0.951&0.953&&0.955& 0.952&0.958&0.953&0.954& 0.944&0.950\\
BCP &0.945& 0.951&0.947&0.924& 0.957&0.952&0.945&&0.953& 0.950&0.951&0.953&0.952& 0.943&0.950\\
\hline
\multicolumn{16}{c}{$n=1,000$}\\
\hline
Bias&0.042&-0.005&-0.008&0.022&-0.002&0.004&0.022&&0.010&0.007&0.003&0.013&0.014&-0.001&0.015\\
RMSE&0.368& 0.354& 0.155&0.091& 0.272&0.293&0.163&&0.091&0.107&0.054&0.077&0.141& 0.185&0.121\\
ASE &0.359& 0.343& 0.157&0.089& 0.273&0.295&0.166&&0.089&0.106&0.057&0.074&0.143& 0.185&0.122\\
BSE &0.366& 0.354& 0.155&0.088& 0.272&0.293&0.162&&0.090&0.107&0.054&0.076&0.140& 0.185&0.120\\
ACP &0.928& 0.949& 0.946&0.957& 0.953&0.957&0.956&&0.945&0.946&0.967&0.942&0.957& 0.957&0.955\\
BCP &0.939& 0.951& 0.950&0.945& 0.950&0.954&0.949&&0.945&0.946&0.955&0.943&0.952& 0.949&0.948\\
\hline
\end{tabular}
\end{table}

\begin{table}[H]
\setlength{\tabcolsep}{0.10cm}
\setstretch{1.1}
\centering
\sf
\scriptsize
\caption{Monte Carlo simulation results for the Gamma-Multinomial regression model.
}
\label{SimuTab.Gamma}
\vspace{-0.2cm}
\begin{tabular}{rrrrrrrrrrrrrrrr}
\hline
\multicolumn{16}{c}{\text{\bf Scenario 1}}\\
\hline
\multirow{2}{*}{} & \multicolumn{7}{c}{Cause 1}& \multicolumn{7}{c}{Cause 2}\\
\cline{2-8} \cline{10-16}
& $\beta_{01}$ & $\beta_{11}$ & $\beta_{21}$ & $\alpha_{1}$ & $\xi_{01}$ & $\xi_{11}$ & $\xi_{21}$ && $\beta_{02}$ & $\beta_{12}$ & $\beta_{22}$ & $\alpha_{2}$ & $\xi_{02}$ & $\xi_{12}$ & $\xi_{22}$ \\
\hline
\multicolumn{16}{c}{$n=200$}\\
\hline
Bias&0.007&0.003&0.005&0.044&-0.012&0.011&0.042&&-0.044&0.030&-0.009&0.075&-0.044&0.057&-0.046\\
RMSE&0.679&0.724&0.427&0.126& 0.384&0.458&0.322&& 0.413&0.458& 0.277&0.206& 0.365&0.469& 0.353\\
ASE &0.637&0.659&0.397&0.117& 0.379&0.462&0.307&& 0.396&0.449& 0.261&0.180& 0.356&0.467& 0.332\\
BSE &0.680&0.725&0.428&0.118& 0.384&0.458&0.318&& 0.410&0.456& 0.276&0.191& 0.361&0.465& 0.349\\
ACP &0.926&0.913&0.926&0.950& 0.959&0.956&0.964&& 0.938&0.945& 0.927&0.951& 0.956&0.955& 0.952\\
BCP &0.946&0.945&0.951&0.927& 0.951&0.945&0.943&& 0.934&0.951& 0.943&0.926& 0.942&0.950& 0.954\\
\hline
\multicolumn{16}{c}{$n=500$}\\
\hline
Bias&-0.017&0.003&0.003&0.014&-0.009&0.014&0.003&&-0.004&-0.004&-0.005&0.025&-0.010& 0.000&-0.017\\
RMSE& 0.424&0.429&0.259&0.075& 0.234&0.290&0.178&& 0.238& 0.285& 0.167&0.110& 0.213& 0.278& 0.200\\
ASE & 0.402&0.417&0.250&0.072& 0.229&0.281&0.184&& 0.249& 0.283& 0.165&0.110& 0.217& 0.286& 0.202 \\
BSE & 0.424&0.428&0.259&0.073& 0.234&0.290&0.177&& 0.239& 0.284& 0.168&0.106& 0.212& 0.277& 0.199\\
ACP & 0.930&0.943&0.938&0.943& 0.954&0.949&0.961&& 0.954& 0.952& 0.944&0.958& 0.951& 0.954& 0.956\\
BCP & 0.945&0.945&0.953&0.943& 0.953&0.955&0.943&& 0.956& 0.952& 0.950&0.944& 0.938& 0.948& 0.945\\
\hline
\multicolumn{16}{c}{$n=800$}\\
\hline
Bias& 0.005&0.002&-0.005&0.008&-0.007&-0.002&0.011&&-0.002&0.001&-0.003&0.017&-0.002&0.005&-0.007 \\
RMSE& 0.316&0.327& 0.199&0.057& 0.178& 0.226&0.143&& 0.200&0.222& 0.135&0.092& 0.170&0.225& 0.158\\
ASE & 0.317&0.328& 0.197&0.057& 0.180& 0.221&0.145&& 0.196&0.223& 0.129&0.086& 0.170&0.225& 0.158\\
BSE & 0.316&0.327& 0.199&0.056& 0.178& 0.226&0.143&& 0.200&0.222& 0.135&0.091& 0.171&0.232& 0.158\\
ACP & 0.949&0.952& 0.952&0.951& 0.956& 0.951&0.951&& 0.947&0.957& 0.944&0.945& 0.949&0.942& 0.958\\
BCP & 0.944&0.948& 0.951&0.952& 0.945& 0.947&0.946&& 0.949&0.955& 0.956&0.948& 0.941&0.953& 0.958\\
\hline
\multicolumn{16}{c}{$n=1,000$}\\
\hline
Bias&-0.004&0.003&-0.001&0.009&0.000&-0.002&0.001&&0.000&-0.009&-0.002&0.013&-0.006&0.003&-0.018\\
RMSE& 0.283&0.288& 0.177&0.051&0.158& 0.196&0.130&&0.180& 0.197& 0.123&0.077& 0.152&0.206& 0.141\\
ASE & 0.282&0.291& 0.175&0.051&0.160& 0.196&0.129&&0.176& 0.200& 0.116&0.077& 0.152&0.201& 0.141 \\
BSE & 0.283&0.288& 0.177&0.050&0.158& 0.196&0.130&&0.180& 0.197& 0.123&0.076& 0.152&0.206& 0.138\\
ACP & 0.940&0.955& 0.946&0.945&0.958& 0.946&0.949&&0.941& 0.949& 0.933&0.951& 0.959&0.948& 0.950\\
BCP & 0.945&0.950& 0.948&0.945&0.943& 0.945&0.947&&0.946& 0.950& 0.948&0.942& 0.950&0.952& 0.935\\
\hline
\multicolumn{16}{c}{\text{\bf Scenario 2}}\\
\hline
\multirow{2}{*}{} & \multicolumn{7}{c}{Cause 1}& \multicolumn{7}{c}{Cause 2}\\
\cline{2-8} \cline{10-16}
& $\beta_{01}$ & $\beta_{11}$ & $\beta_{21}$ & $\alpha_{1}$ & $\xi_{01}$ & $\xi_{11}$ & $\xi_{21}$ && $\beta_{02}$ & $\beta_{12}$ & $\beta_{22}$ & $\alpha_{2}$ & $\xi_{02}$ & $\xi_{12}$ & $\xi_{22}$ \\
\hline
\multicolumn{16}{c}{$n=200$}\\
\hline
Bias&-0.247&0.146&0.006&0.251&-0.045&0.049&-0.078&&-0.020&-0.032&-0.018&0.031&0.121& 0.073&-0.110\\
RMSE& 0.979&1.017&0.473&0.472& 0.923&0.927& 0.549&& 0.375& 0.593& 0.316&0.105&0.554& 0.654& 0.494 \\
ASE & 0.809&0.858&0.385&0.336& 0.873&0.910& 0.486&& 0.395& 0.595& 0.320&0.101&0.549& 0.663& 0.427\\
BSE & 0.947&1.007&0.473&0.400& 0.923&0.926& 0.543&& 0.374& 0.591& 0.314&0.100&0.540& 0.649& 0.482\\
ACP & 0.856&0.889&0.877&0.922& 0.951&0.964& 0.935&& 0.954& 0.948& 0.954&0.958&0.954& 0.972& 0.947\\
BCP & 0.938&0.950&0.948&0.906& 0.938&0.944& 0.945&& 0.954& 0.938& 0.952&0.941&0.939& 0.942& 0.940\\
\hline
\multicolumn{16}{c}{$n=500$}\\
\hline
Bias&-0.020&0.008&0.009&0.071&0.003&0.016&-0.024&&-0.005&-0.012&-0.009&0.015&0.050&-0.028&-0.041\\
RMSE& 0.606&0.622&0.254&0.211&0.569&0.556& 0.309&& 0.252& 0.377& 0.205&0.064&0.330& 0.372& 0.261\\
ASE & 0.582&0.593&0.245&0.189&0.561&0.564& 0.299&& 0.250& 0.374& 0.201&0.063&0.329& 0.391& 0.258\\
BSE & 0.605&0.622&0.254&0.198&0.569&0.556& 0.307&& 0.252& 0.376& 0.205&0.063&0.325& 0.370& 0.257\\
ACP & 0.924&0.934&0.927&0.946&0.951&0.957& 0.946&& 0.958& 0.942& 0.949&0.957&0.962& 0.984& 0.951\\
BCP & 0.948&0.949&0.947&0.933&0.937&0.940& 0.946&& 0.954& 0.946& 0.954&0.950&0.956& 0.959& 0.941\\
\hline
\multicolumn{16}{c}{$n=800$}\\
\hline
Bias&-0.028&0.010&-0.004&0.045&-0.011&0.018&-0.012&&-0.004&-0.002&-0.007&0.006&0.025&-0.008&-0.019\\
RMSE& 0.461&0.477& 0.196&0.154& 0.420&0.430& 0.228&& 0.204& 0.304& 0.161&0.049&0.243& 0.306& 0.193\\
ASE & 0.452&0.462& 0.193&0.146& 0.422&0.428& 0.230&& 0.198& 0.297& 0.159&0.049&0.243& 0.296& 0.196\\
BSE & 0.460&0.477& 0.195&0.148& 0.420&0.430& 0.228&& 0.203& 0.303& 0.161&0.049&0.242& 0.305& 0.192\\
ACP & 0.948&0.948& 0.949&0.952& 0.954&0.949& 0.951&& 0.942& 0.946& 0.939&0.958&0.946& 0.956& 0.959\\
BCP & 0.949&0.948& 0.943&0.924& 0.952&0.951& 0.945&& 0.948& 0.947& 0.952&0.946&0.947& 0.948& 0.942\\
\hline
\multicolumn{16}{c}{$n=1,000$}\\
\hline
Bias&-0.030&0.020&0.000&0.040&-0.004&0.016&-0.021&&0.003&0.006&-0.004&0.007&0.025&-0.004&-0.026\\
RMSE& 0.416&0.435&0.171&0.137& 0.385&0.383& 0.209&&0.180&0.269& 0.143&0.045&0.212& 0.256& 0.179\\
ASE & 0.404&0.413&0.171&0.130& 0.376&0.382& 0.206&&0.177&0.264& 0.142&0.044&0.216& 0.263& 0.175\\
BSE & 0.415&0.434&0.171&0.131& 0.385&0.383& 0.207&&0.179&0.269& 0.143&0.044&0.210& 0.256& 0.177\\
ACP & 0.931&0.935&0.952&0.948& 0.952&0.956& 0.944&&0.946&0.949& 0.942&0.945&0.963& 0.969& 0.952\\
BCP & 0.946&0.949&0.945&0.948&0.944&0.946& 0.943&&0.946&0.952& 0.941&0.946&0.949& 0.947& 0.948\\
\hline
\end{tabular}
\end{table}

\section{Additional results of real data application }\label{AdppA2}
\renewcommand{\thefigure}{A2.\arabic{figure}}

This section presents supplementary results from applying the proposed methodology to the melanoma dataset. These findings complement those in the main text and provide additional insight into model fit and the associations among patient characteristics, cause-specific susceptibility, and event-time distributions.

\begin{figure}[H]
\centering
\begin{tabular}{cc}
\includegraphics[width=0.48\textwidth]{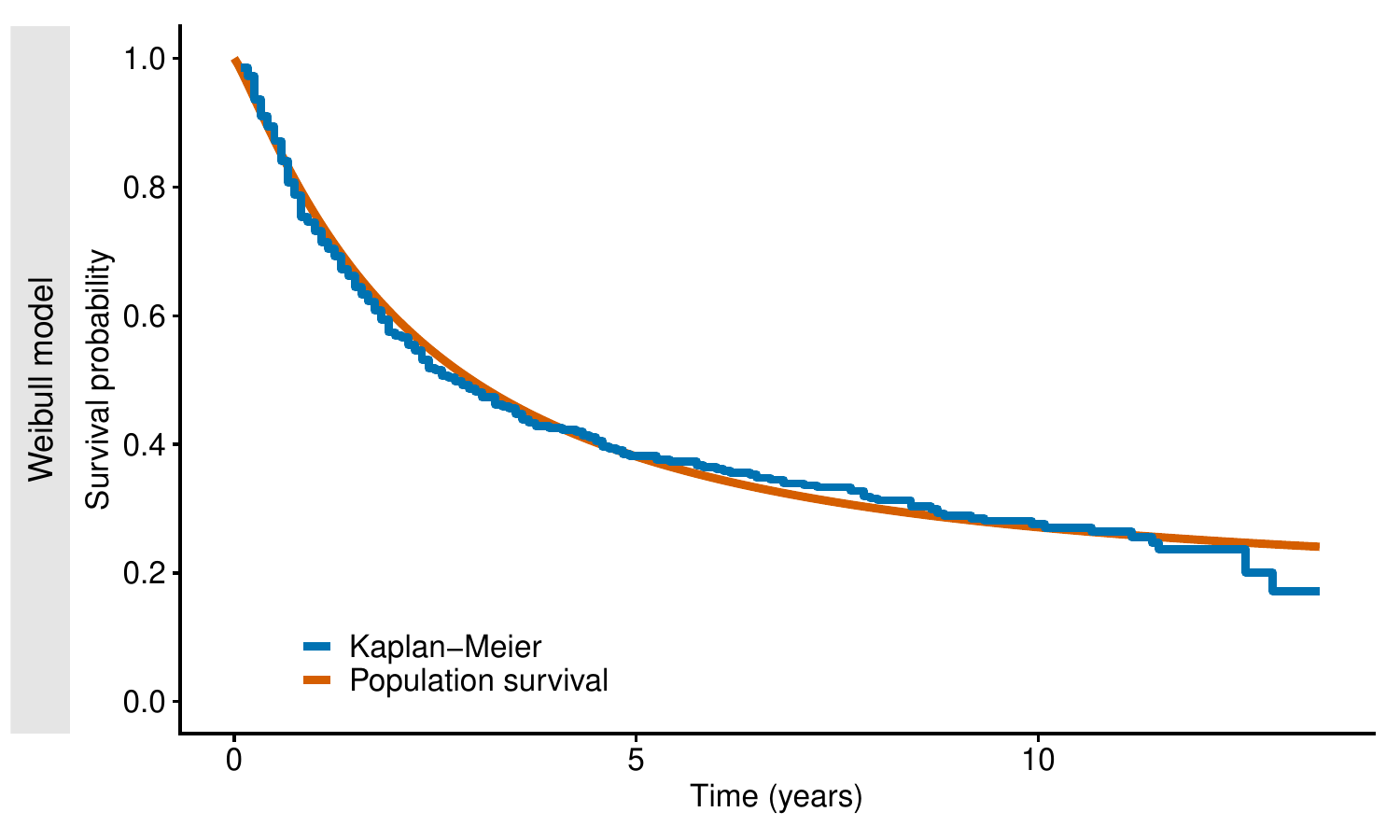} &
\includegraphics[width=0.48\textwidth]{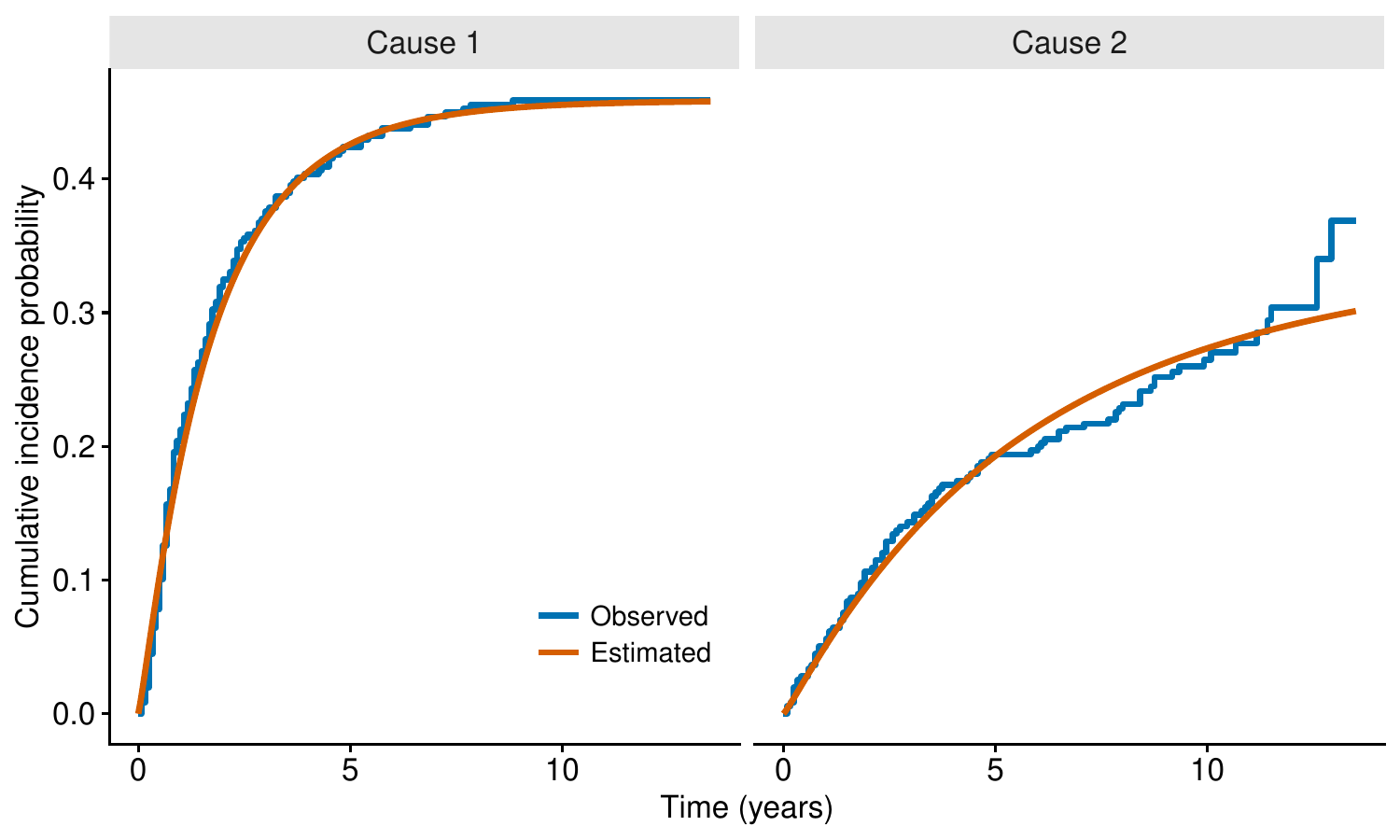} \\[0.2em]

\includegraphics[width=0.48\textwidth]{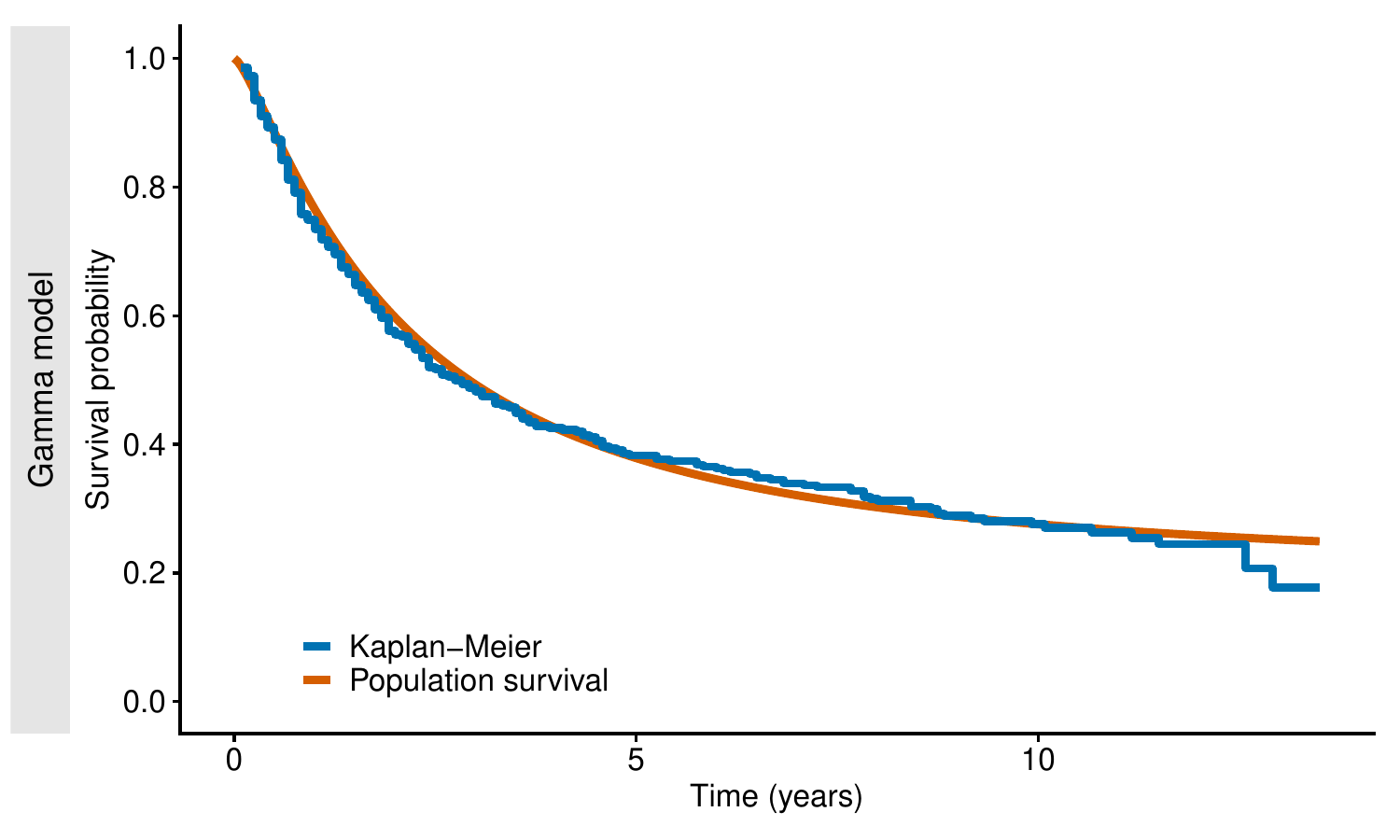} &
\includegraphics[width=0.48\textwidth]{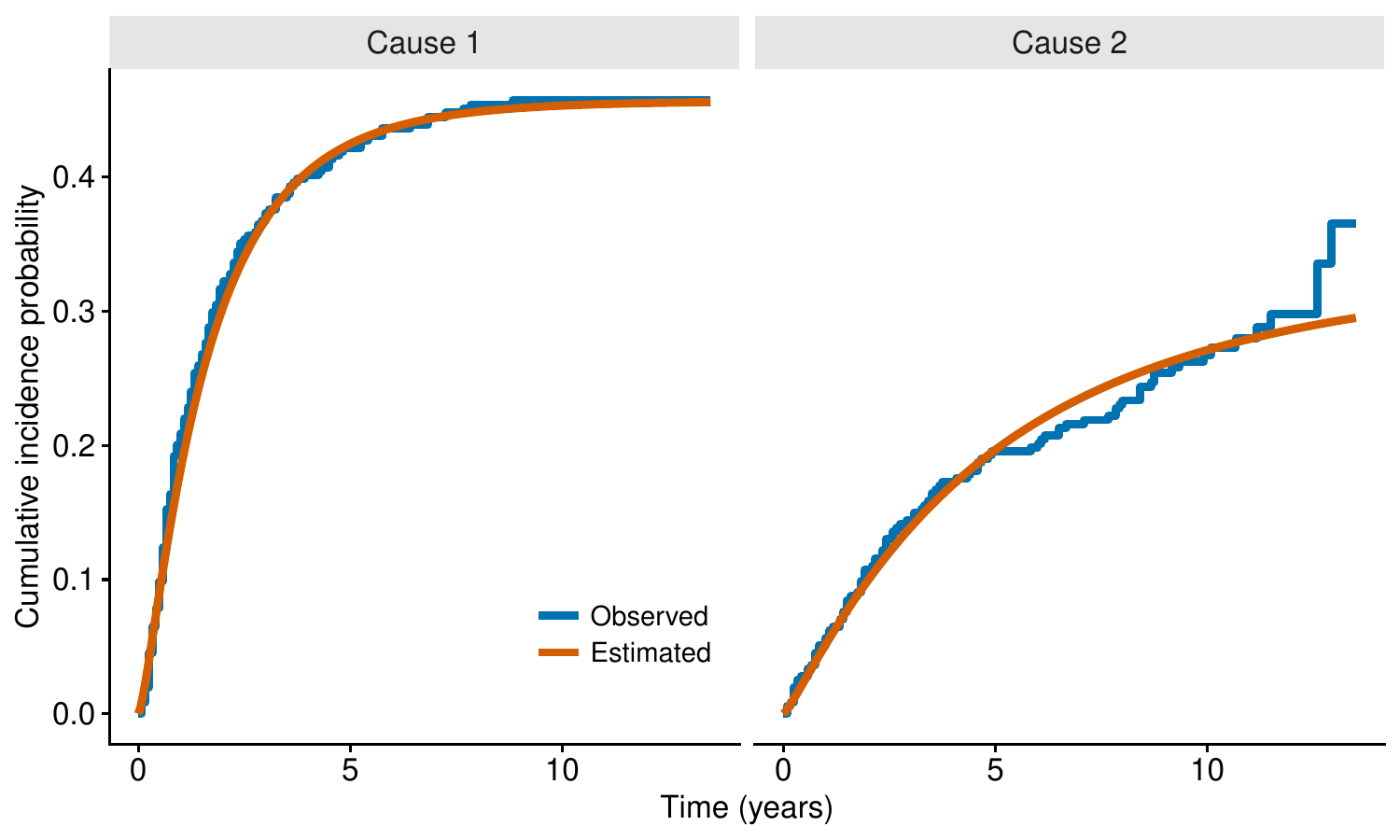} \\[0.2em]

\includegraphics[width=0.48\textwidth]{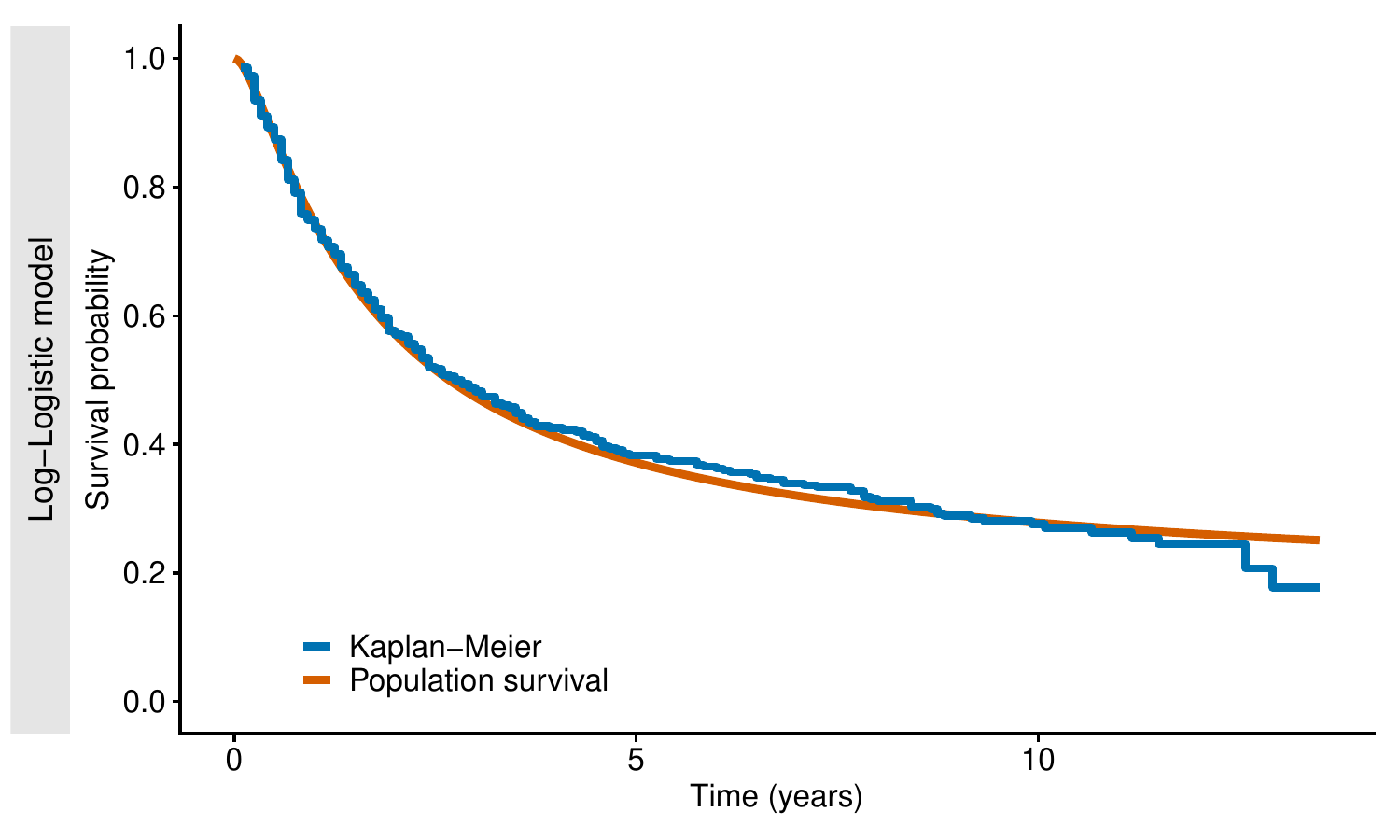} &
\includegraphics[width=0.48\textwidth]{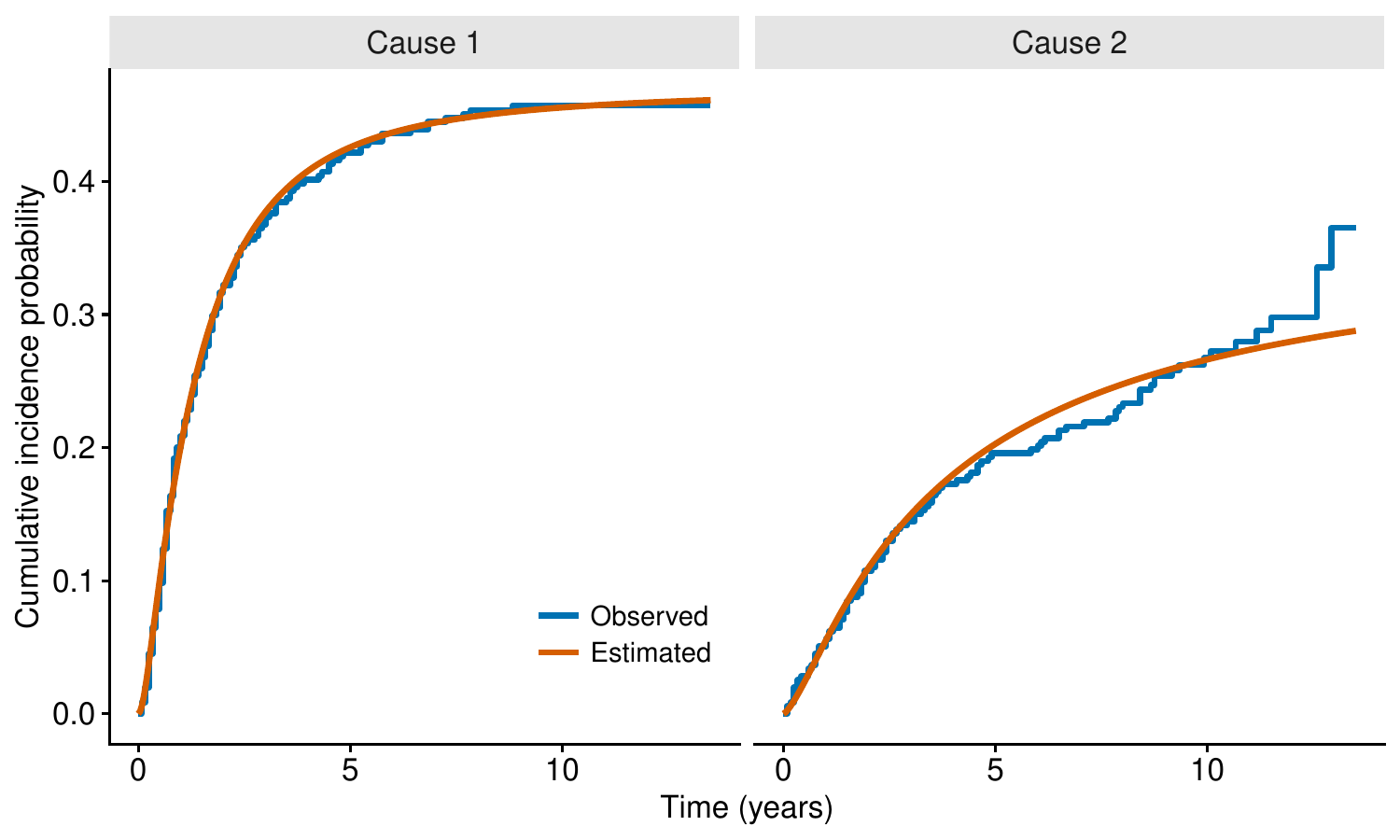}
\end{tabular}
\vspace{-0.2cm}
\caption{Comparison of the fitted survival and cumulative incidence functions for the Weibull, Gamma, and log-logistic mixture cure rate models with competing risks.}
\label{FigModelchoose}
\end{figure}

\end{document}